\documentclass[journal]{IEEEtran}
\input{decl}

\begin{document}

\title{Time encoding of bandlimited signals: reconstruction by pseudo-inversion\\and time-varying multiplierless FIR filtering}

\author{Nguyen T. Thao,~\IEEEmembership{Member,~IEEE,} and
Dominik Rzepka% <-this % stops a space
\thanks{N. T. Thao is with the Department of Electrical Engineering, The City College of New York, CUNY, New York, USA, email: tnguyen@ccny.cuny.edu.}% <-this % stops a space
\thanks{D. Rzepka is with the Department of Measurement and Electronics, AGH University of Science and Technology, Krak\'ow, Poland, email: drzepka@agh.edu.pl.}% <-this % stops a space
\thanks{D. Rzepka was supported by the Polish National Center of Science under Grant DEC-2017/27/B/ST7/03082.}
}

\maketitle

\begin{abstract}
We propose an entirely redesigned framework of bandlimited signal reconstruction for the time encoding machine (TEM) introduced by Lazar and T\'oth. As the encoding part of TEM consists in obtaining integral values of a bandlimited input over known time intervals, it theoretically amounts to applying a known linear operator on the input. We then approach the general question of signal reconstruction by pseudo-inversion of this operator. We perform this task numerically and iteratively using projections onto convex sets (POCS). The algorithm can be implemented exactly in discrete time with multiplications that are all reduced to scaling by signed powers of two, thanks to the use of relaxation coefficients. Meanwhile, the algorithm achieves a rate of convergence similar to that of Lazar and T\'oth. For real-time processing, we propose an approximate time-varying FIR implementation, which avoids the splitting of the input into blocks. We finally propose some preliminary semi-convergence analysis of the algorithm under data noise.
\end{abstract}


\section{Introduction}

\subsection{Context and goal}

\begin{figure}
\centerline{\scalebox{0.8}{\includegraphics{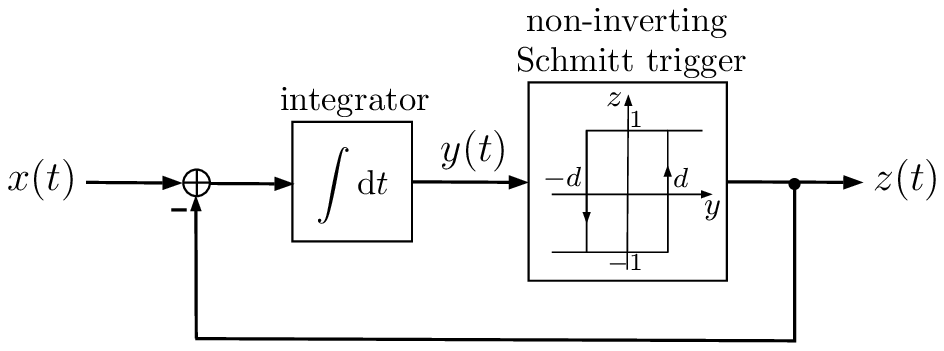}}}
\vspace{4mm}
\centerline{\scalebox{0.7}{\includegraphics{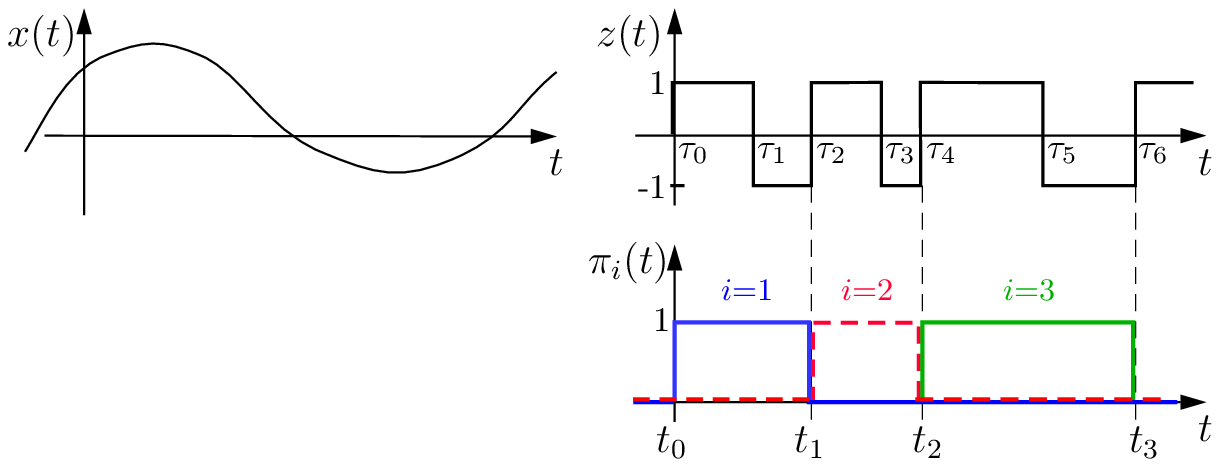}}}
\caption{\label{fig:encoder} Encoder of time encoding machine: asynchronous Sigma-Delta modulator (ASDM)}
\end{figure}

Since its origin, analog-to-digital conversion has been mostly based on uniformly sampling a continuous-time bandlimited signal $x(t)$ at or above the Nyquist rate, followed by a quantization of the samples in amplitude. One thus obtains a digital description from which $x(t)$ is recovered by a one-step sinc interpolation at the precision of quantization. The idea to extend data acquisition to nonuniform sampling has been theoretically studied for quite some time \cite{Duffin52,Yen56,Feichtinger94,Marvasti01,Aldroubi01} but has attracted relatively low attention in signal processing due to the necessity of complex digital postprocessing. This topic has known revived interest with the recent trend of event-based signal processing \cite{Gontier14,Miskowicz2018}. The main motivations behind this movement has been the higher demand for low power and low complexity acquisition devices, while digital postprocessing is becoming more accessible. On the technical side, a main direction of research has been the extraction of samples by time detection of the input's crossings with fixed amplitude levels, rather than by amplitude measurement of the input at fixed instants \cite{sayiner1996level,allier2003new}. One goal is to boost the overall performance of the acquisition by taking advantage of the inherently higher precision of solid-state circuits in time than in amplitude. A breakthrough in this direction has been the use by Lazar and T\'oth of an asynchronous Sigma-Delta modulator (ASDM) \cite{kikkert1975asynchronous} (see Fig. \ref{fig:encoder}) to extract level-crossings of the input in the integral domain \cite{Lazar04,Koscielnik12}. This method owes its appeal to the high simplicity of the time encoder, together with the built-in robustness to analog circuit imperfections inherited from Sigma-Delta modulation \cite{Pavan17}. This gives the perspective of high-precision low-power acquisition devices, up to remote or offline digital postprocessing. This has given higher motivations to reinvest in the difficult theoretical and practical problem of signal recovery from nonuniform samples. Initial reconstruction methods from time-encoded integrals have been proposed by Lazar and T\'oth in \cite{Lazar04,Lazar08} for bandlimited inputs, and more recently in \cite{Alexandru20,Rudresh20} for inputs with finite rate of innovation. In this paper, we propose to revisit the reconstruction of bandlimited signals in this problem, all the way from theoretical foundations to real time implementations.

\subsection{Pseudo-inversion of nonuniform sampling}\label{sec:framework}

The first part of this paper is to reinterpret the method of \cite{Lazar04} from a higher level perspective of nonuniform sampling, and point some fundamental limitations that we address with a new approach.
The initial step is to formalize at the most abstract level the knowledge provided by the output of an ASDM about its bandlimited input $x(t)$. We show that this knowledge is a sequence of samples of the form
\begin{equation}\label{xi}
\s_i:=\langle f_i,x\rangle,\qquad i\in\Z
\end{equation}
where $\Z\subset\ZZ$ and $\langle\cdot,\cdot\rangle$ is the inner-product of $L^2(\RR)$. The functions $(f_i(t))_{i\in\Z}$ are specifically the bandlimited versions of the rectangular functions $(\pi_i(t))_{i\in\Z}$ shown in Fig. \ref{fig:encoder} delimited by the switching instants $(t_i)_{i\in\Z}$ of the ASDM's output. The reconstruction of $x(t)$ proposed in \cite{Lazar04} can be presented as iterating estimates $x\up{n}(t)$ of the form
\begin{equation}\label{x-rec}
x\up{n}(t)=\smallsum{i\in\Z}\c\up{n}_i\,g_i(t)
\end{equation}
where $(g_i(t))_{i\in\Z}$ is a different family of functions, namely, sinc functions located at the midpoints of the rectangular functions $(\pi_i(t))_{i\in\Z}$. This choice of functions $g_i(t)$ may appear peculiar but was based on an algorithm previously devised in \cite{Feichtinger94} that was proved to converge to $x(t)$ provided that $\Z=\ZZ$ and the switching steps $T_i:=t_i-t_{i-1}$ for $i\in\ZZ$ have an upper bound $T_\m$ smaller than the Nyquist period $T$. However, we indicate in this paper that this sampling condition is not necessary for $x(t)$ to be theoretically and uniquely recoverable from \eqref{xi}. Our first contribution is to propose the alternative choice
\begin{equation}\label{our-gi}
g_i(t):=f_i(t)/\|\pi_i\|^2
\end{equation}
where $\|\cdot\|$ is the $L^2$-norm. Making $g_i(t)$ proportional to $f_i(t)$ in \eqref{x-rec} appears to be a more legitimate choice as \eqref{xi} uniquely characterizes its input $x(t)$ basically when the functions $f_i(t)$ span the whole considered space $\scB$ of bandlimited signals. But the precise features and advantages of our method are as follows.

1. The algorithm we use to obtain estimates of the form \eqref{x-rec} with \eqref{our-gi} is specifically a method of projection onto convex sets (POCS) \cite{Combettes93,Bauschke96}, which is convergent with {\em no condition} whatsoever. Their limit $x_\bs(t)$ is automatically equal to $x(t)$ whenever the samples $(\s_i)_{i\in\Z}$ from \eqref{xi} are uniquely characteristic of $x(t)$.

2. The reconstruction algorithm of \cite{Feichtinger94} used in \cite{Lazar04} not only requires the condition $T_\m<T$, but also depends on mathematics that are specific to $(\pi_i(t))_{i\in\Z}$ as rectangular functions. Meanwhile, not only does the POCS method require no condition in the present application, but it is a generic algorithm of set theoretic estimation with more powerful properties of convergence, and a wider range of configurations. The present algorithm actually works with any family of functions $(\pi_i(t))_{i\in\Z}$ that is orthogonal in $L^2(\RR)$. It can therefore be extended to more general schemes of nonuniform sampling, such as integrate-and-fire with leakage \cite{Feichtinger12}. The flexible use of the POCS method has also been anticipated in the recent and independent work of \cite{Adam19b,Adam20} in multi-channel time encoding, for its ability to deal with multiple systems of equations of the type \eqref{xi}. In the present paper, we use another degree of freedom offered by this method, which is the injection of {\em relaxation coefficients} in the iteration. This simultaneously permits an acceleration of the convergence, and a computational simplification of significant impact for circuit implementations, as will be seen in the practical contributions of the paper.

3. When there exist more than one bandlimited solution $u(t)$ to the system of equations $\s_i=\langle f_i,u\rangle$ with $i\in\Z$, then $x_\bs(t)$ is precisely the solution $u(t)$ that minimizes the $L^2$-norm. This type of reconstruction was previously introduced in \cite[\S III.B.2]{Unser94} in the case of shift-invariant generalized sampling, but also
earlier in \cite{Yen56} in the basic case of bandlimited signals with a finite number of samples. Under theoretical conditions that are at least realized by default when $\Z$ is finite, $x_\bs(t)$ is more precisely the result of {\em pseudo-inversion} of the linear operator
\begin{equation}\label{samp-op}
S:u(t)\in\scB\mapsto(\langle f_i,u\rangle)_{i\in\Z}
\end{equation}
on the sequence $(\s_i)_{i\in\Z}$ \cite{Luenberger69}. This operation is a basic reflex in linear algebra when looking at \eqref{xi} as the linear equation $Sx=(\s_i)_{i\in\Z}$ with possibly many solutions. But the powerful result is that the POCS method of this paper persistently converges to the pseudo-inverse in the more general situation where this equation is inconsistent, due to data noise for example.

4. Theoretical results on perfect reconstruction of bandlimited signals such as Shannon sampling theorem or the result of nonuniform sampling of \cite{Feichtinger94} used in \cite{Lazar04} can only work with $\Z=\ZZ$. Meanwhile, the minimal-norm reconstruction $x_\bs(t)$ is well defined in all cases, including in the practical cases where $\Z$ is systematically finite. Evidently, one cannot theoretically obtain $x_\bs(t)=x(t)$ in this situation. But, under a high enough density of samples inside the window of acquisition, $x_\bs(t)$ is expected to deviate from $x(t)$ only close to the boundaries of this window, in a way similar to the deviations one expects to obtain when truncating the Shannon sampling reconstruction formula. Now, these artifacts are usually ignored in practice as the size of $\Z$ is typically virtually infinite compared to instantaneous processing windows.

\subsection{Sliding-window discrete-time implementation}
\label{sec:synchro}

\begin{figure}
\centerline{\scalebox{0.87}{\includegraphics{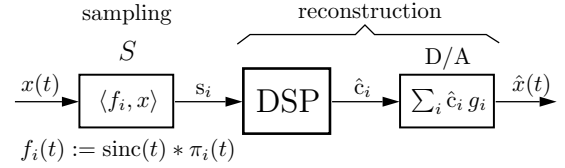}}}
\caption{\label{fig:DSP} Nonuniform but synchronuous DSP of the ASDM output for signal reconstruction}
\end{figure}

The second contribution of this paper is to propose a practical discrete-time implementation of the algorithm. While the POCS method is originally built in the space of continuous-time signals, a pure discrete-time iteration can be obtained by finding a recursive computation of the coefficients $(\c_i\up{n})\subi$ of \eqref{x-rec}. Such a computation was proposed in \cite{Lazar04} involving matrix multiplications of the size of the total signal. Although finite, this size is however virtually infinite compared to the practical windows of operation. For practical implementation, Lazar and T\'oth later abandoned their iterative approach and replaced it by a direct block-based resolution of equation \eqref{xi} \cite{Lazar08}, following some previously developed method in nonuniform point sampling \cite{Strohmer00}. This implies the local algebraic pseudo-inversion of matrices within blocks of signals. As a drawback however, this method creates analytically uncontrolled signal distortions at the block boundaries \cite{Strohmer00}, necessitates ad-hoc and empirical methods of compensations, and remains away from traditional pipeline signal processing.

In this paper, we keep the original goal of a recursive computation of $(\c_i\up{n})\subi$ and propose its approximate implementation by time-varying FIR filtering. Due to the sampling nonuniformity, the matrices of this computation do not have the convolutional (Toeplitz) structure expected in traditional signal processing. However, the intrinsic decay of sinc functions allows to truncate these matrices along subdiagonals that are away enough from the main diagonal, thus permitting time-varying sliding-window processing. While the approach of \cite{Lazar08} is to perform exact algebraic inversions on distorted signals, we return to the traditional signal processing approach, which is on the contrary to preserve the virtually infinite streams of signals while approximating the filtering operators. A similar approach to nonuniform sampling can be found in \cite{Johansson06} in the basic case of point sampling, also with the use of a time-varying FIR reconstruction filter. Contrary to the present paper, this work is non-iterative and involves matrix inversions, but these algebraic manipulations are solely used to approximate the filter. The price to pay however is the inversion of one matrix per sample, of the size of the FIR filter.

With the freedom of relaxation allowed by the POCS method, the discrete-time signals that are processed by our FIR filters can be moreover reduced to mere sequences of signed powers of 2. This permits the replacement of every multiplier of the digital implementation by bit shifters, thus significantly reducing complexity and power consumption \cite{Chandra16}. Paradoxically, this implementation simplification leads to faster convergence than the plain POCS iteration.

\subsection{Nonuniform but synchronous DSP}

Our method is also suggesting a new framework of nonuniform discrete manipulation of bandlimited signals. Typical digital processing from nonuniform samples such as in \cite{Johansson06,Tertinek08} tends to unavoidably reconnect with uniform sampling. This prevents or makes difficult a processing that is adaptive and homogeneous with the local density or the irregularity of the samples. Once minimal-norm reconstruction has been adopted as the goal, all working estimates become limited to the subspace of bandlimited signals
\begin{equation}\label{Vf}
\scV_f:=\overline{\mathrm{span}}(f_i)_{i\in\Z}
\end{equation}
which designates the closed linear span of $(f_i)_{i\in\Z}$, as can be seen in \eqref{x-rec} with \eqref{our-gi}. In this space, every signal has a discrete expansion of the type \eqref{x-rec} (at least when $\Z$ is finite, and more generally when $S$ has a closed range), where the coefficients $(\c_i)\subi$ are in one-to-one correspondence with the nonuniform samples $(\s_i)\subi$. The DSP involved in our reconstruction and symbolized in Fig. \ref{fig:DSP} manipulates these coefficients without any recourse to Nyquist sampling descriptions. In practice, this enables a natural flexible adaptation of the processing with the irregularity of the samples. In theory, this extends DSP to nonuniform discrete descriptions of bandlimited signals.

\subsection{Organization of the paper}

After reviewing the method of \cite{Lazar04} in Section \ref{sec:TEM1} and its limitation, we present our POCS method in Section \ref{sec:POCS}, its unconditional convergence to the minimal-norm bandlimited reconstruction and its connection to pseudo-inversion. In Section \ref{sec:POCS-method}, we introduce relaxation coefficients into the algorithm and show experimental results of reconstruction in comparison with the method of \cite{Lazar04}. In Section \ref{sec:mult-free}, we transform the iterative part of the algorithm into pure discrete-time computation and exploit the relaxation coefficients to make this free of multipliers, assuming that the inner-products $\langle f_i,f_j\rangle$ are available. By neglecting the small terms in this computation, we show in Section \ref{sec:imp} a real-time pipeline implementation of it using time-varying FIR filters, and the consequence of the approximations on the reconstruction results, including time quantization. We then show in Section \ref{sec:lookup-table} how the inner-products $\langle f_i,f_j\rangle$ can be obtained from the sampling steps $(T_i)\subi$ by table lookup and some extra additions. While the overall behavior of the POCS algorithm towards noise is that of pseudo-inversion, we finally give in Section \ref{sec:noise} some early insight on how the error of the iterates moves from the noise-free algorithmic error to the pure noise-induced error.

\section{Time encoding machine}\label{sec:TEM1}

We briefly review the principles of time encoding introduced in \cite{Lazar04} with a particular angle of interest to us. After describing the encoder, we present the reconstruction algorithm of \cite{Lazar04} to eventually point some limitation. All continuous-time signals are assumed to be in the real Hilbert space $L^2(\RR)$ equipped with the inner-product $\langle u,v\rangle:=\int_\RR u(t)v(t)\dif t$ and the norm $\|u\|:=\langle u,u\rangle\upsmall{1/2}$, and $\scB$ is the subspace of bandlimited functions of Nyquist period 1.

\subsection{Encoder}\label{sec:encoder}

The time encoding of a signal $x(t)$ of $\scB$ as proposed in \cite{Lazar04} consists in feeding it into an ASDM as shown in Fig. \ref{fig:encoder} and recording the successive instants $\tau_i$ when the output $z(t)$ switches between $+1$ and $-1$. It is shown in \cite{Lazar04} that
\begin{equation}\label{int-equ}
\int_{ \tau_{i-1}}^{ \tau_i}x(t)\,\dif t=(-1)^i\big(( \tau_i{-} \tau_{i-1})-2d\big)
\end{equation}
where $\pm d$ are the thresholds of the Schmitt trigger. The integral value dependence with the circuit parameter $d$ can be eliminated by considering only the integrals of $x(t)$ between the even-indexed instants $\tau_{2i}$. Defining
\begin{equation}\label{tn-xn}
t_i:= \tau_{2i}\quad\mbox{and}\quad
\s_i:=( \tau_{2i}- \tau_{2i-1})-( \tau_{2i-1}- \tau_{2i-2}),
\end{equation}
one easily obtains from \eqref{int-equ} the relation
\begin{equation}\label{xn}
\s_i=\int_{t_{i-1}}^{t_i}x(t)\,\dif t,\qquad i\in\Z.
\end{equation}
Here, $\Z$ denotes either $\ZZ$ or a finite index set $\{1,\cdots,N\}$. This can be formally rewritten as
\begin{equation}\label{xn2}
\s_i=\langle \pi_i,x\rangle,\qquad i\in\Z
\end{equation}
where
\begin{equation}\label{pii}
\pi_i(t):=1_{[t_{i-1},t_i)}(t)
\end{equation}
and $1_I(t)$ designates the indicator function of any given interval $I$ of $\RR$.
The remaining impact of the parameter $d$ is however in the density of the instants $t_i$.
For certain mathematical orientations, it will be convenient to have an equivalent expression of \eqref{xn2} that involves only bandlimited functions.
Let us define
\begin{equation}\label{fi}
f_i(t):=\sinc(t)*\pi_i(t),\qquad i\in\ZZ
\end{equation}
where $\sinc(t):=\sin(\pi t)/(\pi t)$ and $*$ designates convolution. By even symmetry of the sinc function, we have $\langle \pi_i,{\sinc*u}\rangle=\langle \sinc*\pi_i,u\rangle$. Hence,
\begin{equation}\label{inner-id}
\forall u\in\scB,\qquad\langle \pi_i,u\rangle=\langle f_i,u\rangle.
\end{equation}
Then, \eqref{xn2} is equivalent to \eqref{xi}. Depending on the context, we will preferably refer to \eqref{xi} or to \eqref{xn2}.

\subsection{Generic reconstruction algorithm}\label{sec:gen-alg}

To reconstruct the signal $x=x(t)$ from $\bs=(\s_i)_{i\in\Z}$, \cite{Lazar04} proposes an iteration of the type
\begin{equation}\label{gen-iter}
x\up{n+1}=R_\bs x\up{n}
\end{equation}
where for any $u\in\scB$,
\begin{equation}\label{Rgen}
R_\bs u:=u
+\textstyle\sum\limits_{i\in\Z}\big(\s_i-\langle\pi_i,u\rangle\big)\,g_i
\end{equation}
and $(g_i)_{i\in\Z}$ is some family of $\scB$ to be chosen. Note that any signal $ x_0$ in the solution space
\begin{equation}\label{sol-set}
\scS_\bs:=\big\{u\in\scB:\langle f_i,u\rangle=\s_i,~\forall i\in\Z\big\}
\end{equation}
is a fixed point of \eqref{gen-iter} given the identity of \eqref{inner-id}. One way to make \eqref{gen-iter} convergent is to design $R_\bs$ so that it is a contraction. When this is possible, $\scS_\bs$ is by necessity reduced to a single element, which is $x(t)$, and $x\up{n}(t)$ automatically converges to this signal. In this way, one simultaneously proves uniqueness of reconstruction and provides an algorithm that recovers $x(t)$. It is easy to see that
\begin{equation}\label{linear-part}
R_\bs u-R_\bs v=M(u-v)
\end{equation}
where
\begin{equation}\label{Mu}
Mu:=u-\textstyle\sum\limits_{i\in\Z}\langle\pi_i,u\rangle\,g_i.
\end{equation}
For any $ x_0\in\scS_\bs$, one has
\begin{equation}\label{xn-contract}
x\up{n+1}- x_0=M(x\up{n}- x_0).
\end{equation}
The transformation $R_\bs$ is a contraction when $\|M\|<1$, where $\|\cdot\|$ is here the operator norm in $\scB$.

\subsection{Algorithm configuration of \cite{Lazar04}}

Based on some prior results on frames \cite[$\S8.4$]{Feichtinger94}, the work of \cite{Lazar04} adopts the following functions
\begin{equation}\label{gi}
g_i(t):=\sinc(t-\bar t_i)\quad\mbox{where}\quad \bar t_i:=\smallfrac{1}{2}\,(t_{i-1}+t_i).
\end{equation}
By adapting the derivations of \cite{Feichtinger94}\footnote
{While \cite{Lazar04} assumes that $\bar t_i$ is the midpoint of $(t_{i-1},t_i)$, \cite{Feichtinger94} assumes that $t_i$ is the midpoint of $(\bar t_i,\bar t_{i+1})$ with $T_\m:=\sup_{i\in\Z}\,(\bar t_i-\bar t_{i-1})$.}, it is shown in \cite[Lemma 3]{Lazar04} that $\|M||\leq T_\m$ when $\Z=\ZZ$, where
\begin{equation}\label{Tm}
T_\m:=\sup_{i\in\Z}\,T_i\qquad\mbox{and}\qquad T_i:=t_i-t_{i-1}.
\end{equation}
Thus, $M$ is a contraction when
\begin{equation}\label{lazar-cond}
T_\m<1.
\end{equation}
With the circuit parameters of Fig. \ref{fig:encoder}, it is shown in \cite{Lazar04} that
$T_\m\leq2d/(1{-}x_\m)$ where $x_\m$ is the maximum amplitude of $x(t)$. So $T_\m<1$ is guaranteed by choosing $d<(1{-}x_\m)/2$.

\subsection{Tightness of sampling condition}

Two pending questions arise from the approach of \cite{Lazar04}: (i) $T_\m<1$ is sufficient to make $M$ a contraction, but is this necessary? (ii) making $M$ a contraction is sufficient to guarantee that \eqref{xi} yields $x(t)$ as the unique bandlimited solution, but is this necessary?

An example can be found where the answer is negative for both questions. Consider the case where
\begin{equation}\label{2per-ti}
t_i=i+(-1)^i\delta
\end{equation}
for some constant $\delta\in[0,\half)$. The sampling here is only periodically nonuniform of period 2. Note that the average density of the instants $(t_i)_{i\in\ZZ}$ is 1 while $T_\m=1+2\delta$. By Fourier analysis, we show in Appendix \ref{app:2per} that \eqref{xi} uniquely characterizes $x(t)$ with any $\delta\in[0,\half)$, thus allowing any value of $T_\m<2$. Meanwhile, we find that $\|M\|>1$ as soon as $T_\m>1.72$. This shows that there are cases where uniqueness of reconstruction is realized with $T_\m>1$ while the algorithm of \cite{Lazar04} is not guaranteed to converge. Qualitatively speaking, the iteration \eqref{gen-iter} with the choice of functions $(g_i)_{i\in\Z}$ of \eqref{gi} is not optimally  connected to the property of unique reconstruction of $x(t)$ from \eqref{xi}.

\section{Minimal-norm bandlimited reconstruction}\label{sec:POCS}

We present here the basic principle of our proposed reconstruction method. After characterizing the solution space $\scS_\bs$ and its minimal-norm element $x_\bs(t)$, we show how the POCS method is used to reach this solution. We will then give the more powerful connection of this method to the pseudo-inversion of the sampling operator. This will require some preliminary formalization of sampling from the perspective of operator theory.

\subsection{Set of bandlimited solutions}

Section \ref{sec:gen-alg} gave a sufficient but not necessary condition for \eqref{xi} to uniquely characterize $x(t)$ via the existence of a contractive mapping. As uniqueness of reconstruction lies exactly in the property that $\scS_\bs$ defined in \eqref{sol-set} is a singleton, we wish to get more insight on the structure of this set.
\medskip
\begin{proposition}\label{prop:S}
\begin{equation}\label{SV}
\scS_\bs=x+\scV_f^\perp
\end{equation}
where $\scV_f$ is defined in \eqref{Vf} and $\scV_f^\perp$ is its orthogonal complement in $\scB$.
\end{proposition}
\begin{IEEEproof}
Let $u\in\scB$. Since $\langle f_i,u-x\rangle=\langle f_i,u\rangle-\s_i$, then $u\in\scS_\bs$ if and only if $u-x$ is orthogonal to $f_i$ for all $i\in\Z$. This is in turn equivalent to $u-x\in\scV_f^\perp$. This proves \eqref{SV}.
\end{IEEEproof}
\medskip
Then $\scS_\bs$ is a singleton if and only if $\scV_f^\perp=\{0\}$. Thus, uniqueness of reconstruction is guaranteed only when $\scV_f=\scB$, which means that the family $(f_i)_{i\in\Z}$ spans the whole space $\scB$, qualitatively speaking. This gives an early justification why taking $g_i(t)$ proportional to $f_i(t)$ as introduced in \eqref{our-gi} is a more reasonable choice than \eqref{gi} when estimating $x(t)$ with signals of the form \eqref{x-rec}.

\subsection{Characterization of minimal-norm solution}

Finding the exact condition for $(f_i)_{i\in\Z}$ to span the whole space $\scB$ is a difficult theoretical question that goes beyond the scope of this paper. Our goal is at least to estimate $x(t)$ with some element $x_\bs(t)$ of $S_\bs$. This will guarantee that $x_\bs(t)=x(t)$ whenever $(f_i)_{i\in\Z}$  effectively spans the whole space $\scB$ independently of one's ability to prove it or not. As mentioned in the introduction, the estimate that turns out to be picked by the POCS method is
\begin{equation}\label{x*def}
x_\bs(t):=\argmin_{u\in\scS_\bs}\|u\|.
\end{equation}
Note from \eqref{SV} that $\scS_\bs$ is an affine subspace that is closed (even when $\Z$ is infinite), so its minimal-norm element $x_\bs(t)$ always exists and is unique. Before we proceed to the description of the reconstruction method, here are some outstanding properties of $x_\bs(t)$.
\medskip
\begin{proposition}\label{prop:x*charac}
$x_\bs(t)$ belongs to $\scV_f$. It is moreover
\begin{description}
\item (i) the signal of $\scV_f$ that is closest to any given solution element of $\scS_\bs$ (including $x(t)$),
\item (ii) the unique element of $\scV_f$ that is in the solution space $\scS_\bs$.
    \end{description}
\end{proposition}
\medskip
These are basic results of linear algebra that can be found in \cite{Garcia17} in finite dimension, or in \cite{Christensen08} in the context of bounded operators. The above claims are however valid in any Hilbert space without any assumption on the family $(f_i)_{i\in\Z}$, as elementary consequences of the Pythagorean theorem. We recall their justification in Appendix \ref{app:x*charac}.

\subsection{Minimal-norm reconstruction by POCS}\label{sec:min-norm}

All of the above arguments were solely based on the sampling description of \eqref{xi} without any assumption on $(f_i(t))_{i\in\Z}$. What will make the POCS method an attractive candidate for signal reconstruction is the particular feature from \eqref{fi} that $(f_i(t))_{i\in\Z}$ are the bandlimited versions of $(\pi_i(t))_{i\in\Z}$ which form an {\em orthogonal} family in $L^2(\RR)$. In fact, at the exception of Section \ref{sec:lookup-table} and all experimental results, all the upcoming derivations remain valid with any family $(\pi_i(t))_{i\in\Z}$ that is orthogonal. We recall from the introduction that other applications such as integrate-and-fire with leakage \cite{Feichtinger12} can benefit from this generalization.
Given the identity \eqref{inner-id}, we have the following equivalent description of $\scS_\bs$,
\begin{equation}\label{SPB}
\scS_\bs=\Pi_\bs\cap\scB
\end{equation}
where
$${\Pi_\bs}:=\big\{u\in L^2(\RR):\langle \pi_i,u\rangle=\s_i,~\forall i\in\Z\big\}.$$
As both $\scB$ and ${\Pi_\bs}$ are closed affine subspaces of $L^2(\RR)$ (${\Pi_\bs}$ being the intersection of hyperplanes), the POCS method consists in the following recursion
\begin{equation}\label{x-rec1}
x\up{n+1}=P_\scB P_{\Pi_\bs} x\up{n}
\end{equation}
where $P_\scB$ and $P_{\Pi_\bs}$ are the orthogonal projections onto $\scB$ and ${\Pi_\bs}$, respectively. This converges to the orthogonal projection of the initial estimate $x\up{0}(t)$ onto ${\Pi_\bs}\cap\scB$ \cite{Bauschke96}. When $x\up{0}(t)$ is set as the 0 signal, then $x\up{n}(t)$ tends to the element of $\Pi_\bs\cap\scB$ that is closest to 0 with respect to the $L^2$-norm. This is precisely $x_\bs(t)$ due to \eqref{SPB}. Now, by orthogonality of $(\pi_i(t))_{i\in\Z}$, we have explicitly
\begin{align*}
P_{\Pi_\bs}u&=u+\smallsum{i\in\Z}\big(\s_i-\langle\pi_i,u\rangle\big)\pi_i/\|\pi_i\|^2.
\end{align*}
Thus,
\begin{equation}\label{our-iter}
\forall u\in\scB,\quad P_\scB P_{\Pi_\bs}u=u+\smallsum{i\in\Z}\big(\s_i-\langle\pi_i,u\rangle\big)g_i=R_\bs(u)
\end{equation}
where $R_\bs$ is defined in \eqref{Rgen} with the new functions
\begin{equation}\nonumber%\label{our-gi2}
g_i(t):=P_\scB\pi_i(t)/\|\pi_i\|^2=f_i/\|\pi_i\|^2.
\end{equation}
This is the choice of functions $g_i(t)$ we introduced in \eqref{our-gi}. We assume from {\em now on} this definition of $g_i(t)$. As $x\up{0}(t)=0$, it is clear from \eqref{x-rec1} that $x\up{n}(t)$ remains in $\scB$. We have thus established the following result.
\medskip
\begin{proposition}\label{prop:POCS-conv}
Let $\bs=(\s_i)_{i\in\Z}$ be a sequence such that $\scS_\bs\neq\emptyset$, and $(x\up{n}(t))_{n\geq0}$ be recursively defined by
$$x\up{n+1}=R_\bs x\up{n}$$
with $x\up{0}=0$, where $R_\bs$ is defined in \eqref{Rgen} with the functions $g_i(t)$ of \eqref{our-gi}. Then $x\up{n}(t)$ tends to $x_\bs(t)$.
\end{proposition}
\ppnoi
This time, the convergence of $x\up{n}(t)$ is systematic without any condition ($T_\m$ can be for example infinite). Meanwhile, as a composition of orthogonal projections, $R_\bs$ is {\em a priori} not contractive and only non-expansive.
With $x\up{0}(t)=0$, note from \eqref{SPB}, \eqref{our-iter} and \eqref{our-gi} that $x\up{n}(t)$ actually remains in $\scV_f$ for all $n\geq0$.

\subsection{Operator formalism}

In the introduction, we alternatively presented the minimum-norm reconstruction $x_\bs(t)$ as a result of pseudo-inversion of an operator. This however requires some rigorous mathematical construction that we perform in this section. Consider the space of sequences
$$\textstyle\scD:=\left\{\bc=(\c_i)_{i\in\Z}:\sum_{i\in\Z}(\c_i/\|\pi_i\|)^2<\infty\right\}.$$
This is a Hilbert space equipped with the inner-product
$$\textstyle\langle\bc,\bc'\rangle\subsmall{\scD}:=\sum_{i\in\Z}\c_i\c'_i/\|\pi_i\|^2$$
and the induced norm $\|\bc\|\subsmall{\scD}:=\langle\bc,\bc\rangle\subsmall{\scD}^{1\!/2}$. Then, the mapping $S$ of \eqref{samp-op} is rigorously a linear operator of Hilbert spaces
\begin{equation}\label{S}
S:\begin{array}[t]{ccl}
\scB & \rightarrow & \scD\\
u & \mapsto & \big(\langle f_i,u\rangle\big)_{i\in\Z}
\end{array}.
\end{equation}
Indeed, due to the identity \eqref{inner-id} and the orthonormality of $(\pi_i/\|\pi_i\|)_{i\in\Z}$, we have for all $u\in\scB$,
\begin{equation}\label{hS-norm}
\smallsum{i\in\Z}\big(\langle f_i,u\rangle/\|\pi_i\|\big)^2=
\smallsum{i\in\Z}\big\langle\pi_i/\|\pi_i\|,u\big\rangle^2\leq\|u\|^2
\end{equation}
by Bessel's inequality, which implies that $Su\in\scD$. At this occasion, we are finding that $ S$ is a bounded operator of norm $\| S\|\leq1$. With operator notation, equations \eqref{xi} and \eqref{sol-set} then take the form
\begin{equation}\label{op-equ}
\bs=Sx\qquad\mbox{and}\qquad\scS_\bs=S^{-1}(\bs).
\end{equation}
Let us now define the reconstruction operator
\begin{equation}\label{S*}
S^*:\begin{array}[t]{rcl}
\scD & \rightarrow & \scB\\
\bc & \mapsto & \sum_{i\in\Z}\c_i\,g_i
\end{array}.
\end{equation}
We use the notation $S^*$ because this is precisely the adjoint of $S$ given the definition \eqref{our-gi} of $g_i(t)$. This is seen as follows. For any $u\in\scB$ and $\bc\in\scD$, we have
\begin{align*}
\langle Su,\bc\rangle\subsmall{\scD}&=
\textstyle\sum_{i\in\Z}\langle f_i,u\rangle\c_i/\|\pi_i\|^2
=\left\langle\sum_{i\in\Z}\c_i g_i,u\right\rangle
\end{align*}
using the linearity of $\langle\cdot,\cdot\rangle$ with respect to its first argument and \eqref{our-gi}. Thus $\langle Su,\bc\rangle\subsmall{\scD}=\langle u,S^*\bc\rangle$ according to \eqref{S*}, which proves that $S^*$ is indeed the adjoint of $S$.

With \eqref{inner-id}, the transformations $R_\bs$ and $M$ of \eqref{Rgen} and \eqref{Mu} then yield for $u\in\scB$ the expressions
\begin{equation}\label{op-expr}
R_\bs u=u+S^*(\bs-Su)\quad\mbox{and}\quad Mu=u-S^{*\!}Su.
\end{equation}

\subsection{Pseudo-inversion of sampling operator}\label{sec:samp-op}

Given the equation $\bs=Sx$ of \eqref{op-equ}, one naturally thinks of invoking the pseudo-inverse $S^\dagger$ of $S$ to estimate $x(t)$ from $\bs$. In standard mathematics, this operator exists whenever the range of $S$, denoted by $\ran(S)$, is closed \cite{Luenberger69}. It is defined as
\begin{eqnarray*}
&S^\dagger\bc:=\argmin\limits_{u\in\scM_\bc}\|u\|,\quad\forall\bc\in\scD,\\
\mbox{where}&&\qquad\quad\\
&\scM_\bc:=\big\{u\in\scB:\|S u-\bc\|\subsmall{\scD} \mbox{ is minimized}\big\}.
\end{eqnarray*}
As $\bs\in\ran(S)$, it is easy to see that $\scM_\bs=\scS_\bs$ (the minimum of $\|S u{-}\bs\|\subsmall{\scD}$ being 0) and hence
\begin{equation}\label{S+s}
x_\bs(t)=S^\dagger\bs
\end{equation}
from \eqref{x*def}.
But the full action of the pseudo-inverse is when the sample sequence $\bs$ is corrupted by noise. Assume that one only has access to
\begin{equation}\label{noise}
\hbs=\bs+\bn
\end{equation}
where $\bn$ is some noise sequence. One is left with the corrupted POCS iteration
\begin{equation}\label{x-rec2}
x\up{n+1}=R_\hbs\,x\up{n}.
\end{equation}
Because $\hbs$ may no longer be in $\ran(S)$, then $\scS_\hbs$ may be empty and hence $x_\hbs(t)$ may not exist. The following result will help analyze this iteration.
\medskip
\begin{proposition}\label{prop:pseudo}
Under the condition that $\ran(S)$ is closed, let $\bbs$ be the orthogonal projection of $\hbs$ onto $\ran(S)$ with respect to the inner-product $\langle\cdot,\cdot\rangle\subsmall{\scD}$.
Then
\begin{equation}\nonumber
\scM_\hbs=\scM_\bbs\qquad\mbox{and}\qquad R_\hbs=R_\bbs.
\end{equation}
\end{proposition}

\begin{IEEEproof}
For all $u\in\scB$, $Su-\hbs=(Su-\bbs)+(\bbs-\hbs)$ where the last two terms belong to $\ran(S)$ and $\ran(S)^\perp$, respectively. So by the Pythagorean theorem,
$\|Su-\hbs\|\subsmall{\scD}^2=\|Su-\bbs\|\subsmall{\scD}^2+{\|\bbs-\hbs\|\subsmall{\scD}^2}$. As the last term does not depend on $u$, then $\scM_\hbs=\scM_\bbs$. Next, it is easy to see from \eqref{op-expr} that $R_\hbs u=R_\bbs u+{S^*(\hbs-\bbs)}$. But $\hbs-\bbs$ is in $\ran(S)^\perp$ which is equal to the null space of $S^*$ \cite[\S6.6]{Luenberger69}. So $S^*(\hbs-\bbs)=0$.
\end{IEEEproof}
\ppnoi
It follows from this proposition that $x\up{n+1}=R_\bbs\,x\up{n}$ for all $n\geq0$. This time, $\bbs\in\ran(S)$. So
\begin{equation}\label{noise-lim}
\lim_{n\rightarrow\infty}x\up{n}(t)=x_\bbs(t).
\end{equation}
Now, like in \eqref{S+s}, $x_\bbs(t)=S^\dagger\bbs$. But since $\scM_\bbs=\scM_\hbs$ according to the above proposition, we have $S^\dagger\bbs=S^\dagger\hbs$. We have thus established the following result.
\medskip
\begin{proposition}
Assume that $S$ has a closed range. For any given $\hbs\in\scD$, the iterates $x\up{n}(t)$ recursively defined by \eqref{x-rec2} with $x\up{0}(t)=0$ tend to $S^\dagger\hbs$.
\end{proposition}
\pp
The issue of closed range of $S$ is often not raised in engineering publications as this property is automatically realized in finite dimension. This is the case in this paper as soon as $\Z$ is finite, which is always true in practice. The case of an infinite set $\Z$ is mostly of interest for general theorems of harmonic analysis. With the specific functions $(f_i(t))_{i\in\Z}$ of \eqref{fi}, it can be at least currently claimed that $\ran(S)$ is closed when $T_\m<1$ based on the knowledge established in \cite{Feichtinger94} that $(f_i(t)/\|\pi_i\|)_{i\in\ZZ}$ is a {\em frame}. We show in Appendix \ref{app:surjective} that this is also achieved with arbitrarily large $T_\m\geq1$ when the sampling-step sequence $(T_i)\subi$ is 2-periodic with an average that is no less than 1. To the best of the authors' knowledge, this question has not been approached yet by the mathematical community of harmonic analysis.

\section{POCS with relaxation coefficients}\label{sec:POCS-method}

A powerful feature of the POCS method is the possibility to inject relaxation coefficients in its iteration and yet maintain its convergence \cite{Bauschke96}. This feature is used in this paper both to accelerate the convergence and reduce the complexity of implementation as will be seen in Section \ref{sec:mult-free}. The basic principle of relaxation is to replace every orthogonal projection $P$ involved in the iteration by the more general transformation
$P^\lambda u:=u+\lambda(Pu-u)$ for some coefficient $\lambda\in[0,2]$. Note that $P^1=P$. Meanwhile, $P^0u=u$ while $P^2u$ is the mirror of u about $Pu$. With $\lambda\in[0,2]$, $P^\lambda u$ thus takes all positions in the segment $[u,P^2u]$ whose midpoint is $Pu$.
The relaxation method we use in this paper is however somewhat more complex. We successively give the exact description of our relaxed iteration, establish some new resulting facts of convergence, give some insight on the effect of relaxation on convergence rate, and finally give experimental results on this effect with comparisons with the iteration of \cite{Lazar04}.

\subsection{Relaxed iteration}

For any vector of coefficients $\blambda=(\lambda_i)\subi$, we propose to relax the transformation $R_\bs$ of \eqref{Rgen} as
\begin{equation}\label{Rgen-rel}
R_\bs^\blambda u:=u
+\textstyle\sum\limits_{i\in\Z}\lambda_i\big(\s_i-\langle\pi_i,u\rangle\big)\,g_i.
\end{equation}
Then, for a given sequence of vectors $\blambda\up{n}$ in $[0,2]^\Z$, we consider more generally the iteration
\begin{equation}\label{x-rec-rel}
x\up{n+1}=R_\bs^{\blambda\up{n}}x\up{n}.
\end{equation}
To study the convergence of $x\up{n}(t)$, we extract the linear part $M^\blambda$ of $R_\bs^\blambda$.
Similarly to \eqref{linear-part} and \eqref{Mu}, we have $R_\bs^\blambda u-R_\bs^\blambda v=M^\blambda(u-v)$ where
\begin{equation}\label{Mu2}
M^\blambda u:=u-\textstyle\sum\limits_{i\in\Z}\lambda_i\langle\pi_i,u\rangle\,g_i.
\end{equation}
Since $x_\bs$ belongs to $\scS_\bs$, it is a fixed point of $R_\bs^\blambda$ for {\em any} $\blambda$. So, similarly to \eqref{xn-contract},
\begin{equation}\label{rel-contract}
x\up{n+1}-x_\bs=M^{\blambda\up{n}}(x\up{n}-x_\bs).
\end{equation}
One can see from \eqref{Mu2} and \eqref{our-gi} that $M^\blambda$ leaves $\scV_f$ invariant. As $x\up{0}(t)=0$, then $x\up{n}(t)$ remains again in $\scV_f$. It is therefore sufficient to study $M^\blambda$ in $\scV_f$.
We show in Appendix \ref{app:ineq} the following result.
\medskip
\begin{theorem}\label{theo:ineq}
\begin{equation}\nonumber
\forall\blambda\in(0,2)^\Z,~~\forall u\in\scV_f\backslash\{0\},\qquad\|M^\blambda u\|<\|u\|.
\end{equation}
\end{theorem}
This implies a strict decrease of $\|x\up{n}\!-x_\bs\|$ as long as $x\up{n}(t)\neq x_\bs(t)$. This is however not sufficient to imply the convergence of $x\up{n}(t)$ to $x_\bs(t)$. This question is difficult when $\Z$ is infinite. To obtain a firm result  of convergence, we will limit ourselves to the case of interest to us where $\Z$ is finite. It will also be necessary to assume the stronger condition that the relaxation coefficients remain in an interval of the type $[\eps,2{-}\eps]$ for some constant $\eps>0$ as a classic assumption in the literature \cite{Herman75}.
\medskip
\begin{corollary}\label{corol:contraction}
Assume that $\Z$ is finite. For any $\eps\in(0,1]$, there exists a positive constant $\gamma_\eps<1$ such that
\begin{equation}\label{cor-ineq}
\forall\blambda\in[\eps,2{-}\eps]^\Z,~~\forall u\in\scV_f,\quad\|M^\blambda u\|\leq \gamma_\eps\|u\|.
\end{equation}
\end{corollary}

\begin{IEEEproof}
Let $U$ be the unit sphere of $\scV_f$.
Since $\|M^\blambda u\|$ is a continuous function of $(\blambda,u)$ and the set $C:=[\eps,2{-}\eps]^\Z\times U$ is compact, the value
$\gamma_\eps:=\sup_{(\blambda,u)\in C}\|M^\blambda u\|$ is reached at some pair $(\blambda_0,u_0)\in C$. Theorem \ref{theo:ineq} then implies that $\gamma_\eps<\|u_0\|=1$. When $\blambda\in[\eps,2{-}\eps]^\Z$ and $u\in\scV_f\backslash\{0\}$, $\|M^\blambda u\|/\|u\|=\big\|M^\blambda(u/\|u\|)\big\|\leq \gamma_\eps$, which implies \eqref{cor-ineq}.
\end{IEEEproof}
\ppnoi
While $R_\bs$ could not be claimed to be a contraction with an infinite set $\Z$ in Section \ref{sec:min-norm}, its general relaxed version $R_\bs^\blambda$ is seen above to be a contraction within $\scV_f$ when $\Z$ is finite and $\blambda\in[\eps,2{-}\eps]^\Z$.
We then conclude the following result.
\medskip
\begin{proposition}\label{prop:relax}
Assume that $\Z$ is finite. For any given sequence of vectors $(\blambda\up{n})_{n\geq0}$ in $[\eps,2{-}\eps]^\Z$ where $\eps>0$, the iterates $x\up{n}(t)$ of \eqref{x-rec-rel} starting from $x\up{0}(t)=0$ tend to $x_\bs(t)$.
\end{proposition}
\medskip
\begin{IEEEproof}
At each $n\geq0$, we can apply \eqref{cor-ineq} with $\blambda=\blambda\up{n}$. It then follows from \eqref{rel-contract} that
$\|x\up{n+1}\!-x_\bs\|\leq \gamma_\eps\|x\up{n}\!-x_\bs\|$ where $\gamma_\eps<1$ for all $n\geq0$.
\end{IEEEproof}

\subsection{Frame algorithm and over-relaxation}\label{sec:frame}

One wishes to have some insight on the dependence of the convergence rate with the relaxation coefficients. Analytically, this amounts to seeing how small $\gamma_\eps$ can be made in \eqref{cor-ineq} compared to 1, depending on $\blambda$. Its smallest possible value is in fact the operator norm of $M^\blambda$ restricted to $\scV_f$. Let us formally define
$$\|M^\blambda\|:=\inf\limits_{u\in\scV_f\backslash\{0\}}\midfrac{\|M^\blambda u\|}{\|u\|}.$$
As the general analysis of $\|M^\blambda\|$ in terms of $\lambda$ is difficult, one wishes to have at least some idea of this quantity when the components $\lambda_i$ of $\blambda$ are equal to a constant value $\lambda$. For convenience, we will simply write in this case $M^\blambda=M^\lambda$, and the goal is to find
$$\lambda_\m:=\argmin_{\lambda\in\RR}\|M^\lambda\|.$$
In this situation, \eqref{x-rec-rel} coincides with a frame algorithm within $\scV_f$ \cite{Duffin52,Grochenig93} and the optimization of $\|M^\lambda\|$ is classic knowledge. Defining the bounds
$$A:=\inf\limits_{u\in\scV_f\backslash\{0\}}\midfrac{\|Su\|\subsmall{\scD}^2}{\|u\|^2}
\quad\mbox{and}\quad
B:=\sup\limits_{u\in\scV_f\backslash\{0\}}\midfrac{\|Su\|\subsmall{\scD}^2}{\|u\|^2},$$
we have the following result.
\medskip
\begin{proposition}\label{prop:frame-opt}
$\lambda_\m=\frac{2}{A+B}$ and
$\|M^{\lambda_\m}\|=\frac{B-A}{B+A}.$
\end{proposition}
\medskip
As the settings of this paper are not identical to those of \cite{Duffin52,Grochenig93} (e.g., $\scD$ is not $\ell^2(\Z)$) and the present conditions are weaker (e.g., the frame conditions are not guaranteed when $\Z$ is infinite and $T_\m\geq1$), we justify this result in Appendix \ref{app:frame-opt} by adapting the derivations of these references to the present assumptions.

Since $0\leq A\leq B$ with $B\leq 1$ due to \eqref{hS-norm}, then $\lambda_\m\geq1$. In practice, it is likely that $A<B$, which implies that $\lambda_\m>1$. This falls in the case of {\em over-relaxation}, which is typically the result of optimal relaxation with parallel projections \cite{Bauschke06}, but derived here by connection to the frame algorithm. In practice, as $A$ and $B$ may not be analytically available, the value of $\lambda$ is to be  optimized empirically.

\subsection{Experimental results}\label{sec:exp}

\begin{figure}
\includegraphics[width=0.99\columnwidth]{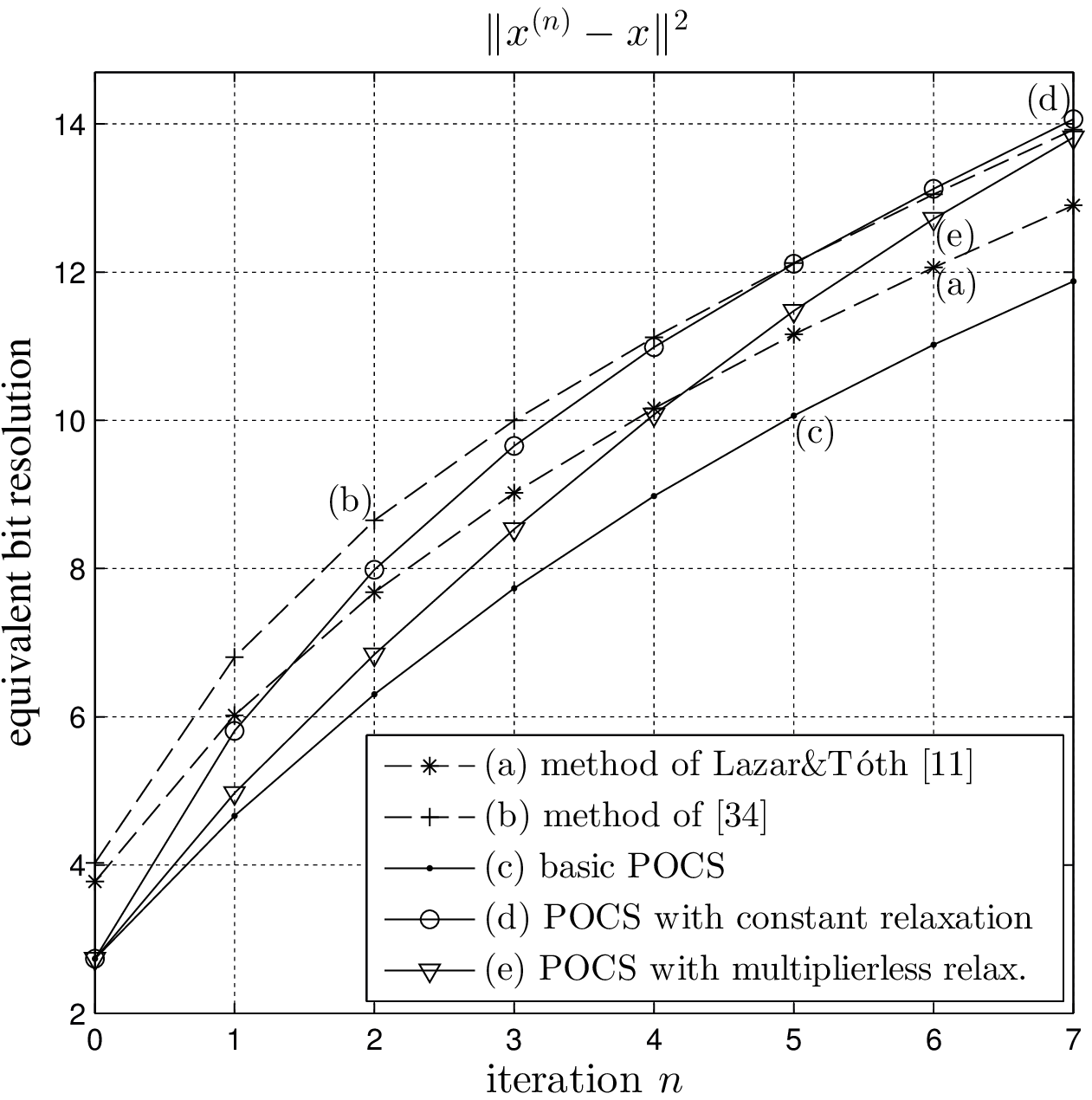}
\caption{MSE of $n$th reconstruction estimate $x\up{n}(t)$ of $x(t)$ of various algorithms from sequence $\{(t_i,\s_i)\}_{i\in\Z}$ of \eqref{tn-xn} with oversampling ratio of 1.5. The results are averaged over 1500 randomly drawn input signals.\label{fig2}}
\vspace{2mm}
\includegraphics[width=0.99\columnwidth]{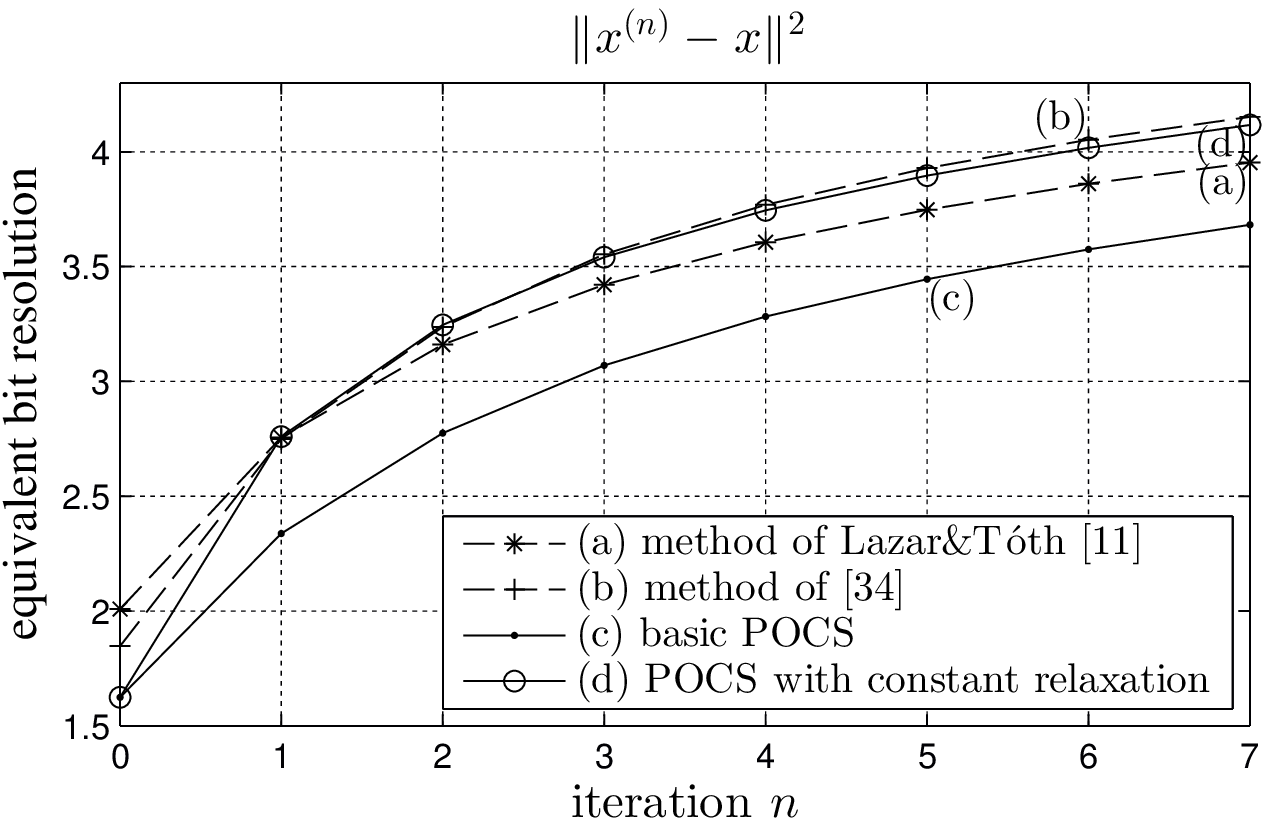}
\caption{Same experiments (a)-(d) as in Fig. \ref{fig2} at critical Nyquist rate.\label{fig:Nyq-rate}}
\end{figure}

We plot in Fig. \ref{fig2} experimental results of mean squared error (MSE) $\|x\up{n}\!-x\|^2$ versus $n$ for various iteration methods, in the case where $\scB$ is the finite dimensional space of bandlimited functions of Nyquist period 1 and signal period $257$. Given the application of the time encoder as an A/D converter, we express the MSE in terms of the bit resolution of a flash A/D converter yielding the same MSE value under the standard uniform quantization noise model \cite{kester2009taking}, with the simple relation 1 bit $= -6.02$ dB of MSE\footnote
{If $b$ is the equivalent bit resolution, the MSE in dB's is equal to $-6.02\,b+\mathrm{MSE}_0$, where MSE$_0$ is the MSE of a random noise uniformly distributed in amplitude and of same maximum amplitude as the signal that is being acquired.}. The samples $\{(t_i,\s_i)\}_{i\in\Z}$ are obtained from the encoding method of Section \ref{sec:encoder} over one period of an input $x(t)\in\scB$ whose Nyquist-rate samples are randomly and uniformly drawn in the amplitude interval $[-0.5,0.5]$. We adjust the parameter $d$ of the Schmitt trigger (indicated in Fig. \ref{fig:encoder}) so that the average density of instants $t_i$ is around 1.5 per Nyquist period. In this situation, $\scS_\bs$ has a unique element and hence $x_\bs(t)=x(t)$. We start with $x\up{0}(t)=0$. The plotted MSE is averaged over 1500 drawn inputs.

As a reference, we show the result of \cite{Lazar04} in (a). The plain POCS method shown in (c) appears to be somewhat inferior. However, the result of (a) is outperformed by the POCS method with a constant relaxation coefficient $\lambda=1.3$ as shown in (d). This constant value has been found by trial and error to give the best result for the class of random inputs described above.
For reference, we also report in (b) the result of another method from \cite[Section V]{Thao16}. This work modifies the sinc functions of \eqref{gi} by deforming their in-band responses to counteract the in-band distortions of the rectangular functions $\pi_i(t)$. This appears to give the best results overall. This technique can however not be applied to the POCS method. Meanwhile, relaxation cannot be used to improve it either (as well as the method of \cite{Lazar04}).

We show in Fig. \ref{fig:Nyq-rate} the same experiments with a value $d$ that yields an average density of instants $t_i$ close to the Nyquist rate (with an optimal value of $\lambda$ equal to 2 in (d)). The figure shows a degradation of all methods with a maximum 4 bit resolution at the 7th iteration while Fig. \ref{fig2} shows 13 bits of resolution. We explain this by the impact of oversampling on the switching regularity of the ASDM (note that a constant input gives perfectly uniform switching instants). Meanwhile, the iterative reconstructions converge faster with more uniform samples. Increasing the sampling rate however also increases the computation cost due to a higher amount of data to be processed. We will keep working with the conditions of Fig. \ref{fig2} in the rest of the paper.

The goal of these experiments has been to show how the POCS method performs compared to the existing methods of \cite{Lazar04,Thao16} and show the potential of relaxation for enhancing its performance. Until now, the relaxation coefficients have been chosen to be constant for reference. But the result of interest to us is the curve (e) of Fig. \ref{fig2}, which uses time-varying coefficients for the sake of implementation simplifications. We present this method in the next section.

\section{Multiplierless digital implementation}\label{sec:mult-free}

The goal of this section is to devise a discrete-time implementation of the relaxed POCS iteration of \eqref{x-rec-rel}. After giving general equations with arbitrary relaxation coefficients, we show how the freedom of relaxation can be used to reduce every multiplication of the iteration to mere bit shifting. This assumes the availability of the inner-products $\langle f_i,f_j\rangle$ whose computation will be presented in Section \ref{sec:lookup-table}. We conclude this section with experimental results of the multiplierless technique.

\subsection{Discrete-time algorithm and continuous-time output}\label{sec:discr-alg}

With \eqref{Rgen-rel} and \eqref{our-gi}, the relaxed POCS iteration of \eqref{x-rec-rel} can be presented as
\begin{equation}\label{x-rec3}
x\up{n+1}(t)=x\up{n}(t)+\smallsum{i\in\Z}\b_i\up{n}f_i(t)
\end{equation}
where
\begin{equation}\label{bi0}
\b_i\up{n}:=\lambda\up{n}_i\big(\s_i-\langle\pi_i,x\up{n}\rangle\big)/\|\pi_i\|^2.
\end{equation}
Thus
\begin{equation}\label{nth-2}
x\up{n}(t)=\smallsum{i\in\Z}\c_i\up{n}f_i(t)
\end{equation}
where the coefficients $\c_i\up{n}$ are recursively obtained from $\b_i\up{n}$ by
\begin{equation}\label{cb}
\c_i\up{n+1}=\c_i\up{n}+\b_i\up{n}
\end{equation}
for all $i\in\Z$, starting from $\c_i\up{0}=0$.
The strategy is to find a pure discrete-time method to obtain the coefficients $\b_i\up{n}$, then output $x\up{n}(t)$ by performing the D/A conversion operation of \eqref{nth-2} only once at the $n$th targeted iteration.
Note that \eqref{nth-2} matches the form of \eqref{x-rec} with \eqref{our-gi} up to some normalization factor $\|\pi_i\|^2$.
Given \eqref{fi}, \eqref{nth-2} yields the alternative expression
\begin{equation}\label{D/A}
x\up{n}(t)=\sinc(t)*\textstyle\sum\limits_{i\in\Z}\c\up{n}_i\,\pi_i(t).
\end{equation}
This is nothing but the bandlimited version of the piecewise constant function equal to $\c\up{n}_i$ in $[t_{i-1},t_i)$ for each $i\in\Z$. This is implemented in circuits by a zero-order hold followed by a lowpass filter. Note that this type of D/A conversion is more suitable to analog circuits than pure sinc reconstructions which ideally require to lowpass filter a Dirac impulse train.
Alternatively, one can extract a PCM description of $x\up{n}(t)$ directly from $\bc\up{n}=(\c_i\up{n})\subi$ by getting from \eqref{nth-2} $x\up{n}(kT)=\sum\subi\c_i\up{n}f_i(kT)$ for some uniform sampling period $T\leq1$.

\subsection{Discrete-time iteration}

We now concentrate on the recursive computation of $\bb\up{n}=(\b_i\up{n})_{i\in\Z}$ needed for \eqref{cb}. To simplify the expression of \eqref{bi0}, note from \eqref{pii} that
$$\|\pi_i\|^2=T_i.$$
While we will apply this result in the remainder of the paper for notation simplification, keep in mind that all the derivations of Sections \ref{sec:mult-free} and \ref{sec:imp} remain valid with any orthogonal family $(\pi_i(t))\subi$ up to changing $T_i$ back to $\|\pi_i\|^2$.
It follows from \eqref{bi0} that
\begin{align}
\b_i\up{n}&\;=\lambda\up{n}_i\r\up{n}_i/T_i\label{bin-res}\\
\mbox{where}\qquad\quad
\r_i\up{n}&:=\s_i-\langle\pi_i,x\up{n}\rangle.\qquad\quad\quad\label{res}
\end{align}
With \eqref{x-rec3}, one finds the recursive relation on $\r_i\up{n}$
$$\r_i\up{n+1}=\r_i\up{n}-\smallsum{j\in\Z}\langle\pi_i,f_j\rangle\,\b_j\up{n}.$$
With the vector notation $\br\up{n}=(\r_i\up{n})_{i\in\Z}$, we obtain the system of equations
\begin{subequations}\label{sys}
\begin{align}
\bb\up{n}&=\big(\lambda\up{n}_i\r\up{n}_i\!/T_i\big)_{i\in\Z}\label{sysa}\\
\br\up{n+1}&=\br\up{n}-\bA\bb\up{n}\label{sysb}
\end{align}
\end{subequations}
where $\bA$ is the matrix of coefficients $\langle\pi_i,f_j\rangle$. With the identity \eqref{inner-id}, we define $\bA$ more elegantly as
\begin{equation}\label{L}
\bA:=\big[\langle f_i,f_j\rangle\big]_{i,j\in\Z}.
\end{equation}
We will see in Section \ref{sec:lookup-table} how the coefficients $\langle f_i,f_j\rangle$ can be obtained using a one-variable lookup table plus a few additions.
Since $x\up{0}=0$, we obtain from \eqref{res} that the initial state of \eqref{sys} is $\br\up{0}=\bs$.

\subsection{Relaxation function}

The next goal is to adjust the coefficients $\lambda\up{n}_i$ so that the global complexity of the system \eqref{sys} is low. Instead of choosing $\b_i\up{n}$ in the form of \eqref{bin-res}, we take
\begin{equation}\label{bi1}
\b\up{n}_i=\beta_i(\r\up{n}_i)
\end{equation}
where $\beta_i(\r)$ is some low complexity function such that $\beta_i(0)=0$. This amounts to \eqref{bin-res} with the time-varying relaxation coefficients
\begin{equation}\label{lambda-beta}
\lambda\up{n}_i:=\left\{\begin{array}{cc}
T_i\,\beta_i(\r\up{n}_i)/\r\up{n}_i, & \r\up{n}_i\neq0\\
1,& \r\up{n}_i=0
\end{array}\right..
\end{equation}
By imposing the function $\beta_i(\r)$ to satisfy the condition
\begin{equation}\label{beta-cond}
\beta_i(0)=0\quad\mbox{and}\quad\forall\r\neq0,\quad T_i\,\beta_i(\r)/\r\in[\eps,2{-}\eps],
\end{equation}
for every $i\in\Z$, we guarantee that the coefficient $\lambda\up{n}_i$ of the equivalent form \eqref{bin-res} remains in $[\eps,2{-}\eps]$ for all $n\geq0$. We thus ensure the convergence of $x\up{n}(t)$ to $ x_\bs(t)$ thanks to Proposition \ref{prop:relax} when $\Z$ is finite.
With \eqref{bi1}, \eqref{sysb} and \eqref{cb}, the vector $\bc\up{n}=(\c_i\up{n})\subi$ is then recursively obtained by the system
\begin{subequations}\label{sys2}
\begin{align}
\bb\up{n}&=\bB(\br\up{n})\label{sys2a}\\
\br\up{n+1}&=\br\up{n}-\bA\bb\up{n}\label{sys2b}\\
\bc\up{n+1}&=\bc\up{n}+\bb\up{n}\label{sys2c}
\end{align}
\end{subequations}
starting with $(\br\up{0},\bc\up{0})=(\bs,\bzero)$, where for any $\br=(\r_i)_{i\in\Z}$,
\begin{equation}\label{B}
\bB(\br):=\big(\beta_i(\r_i)\big)_{i\in\Z}.
\end{equation}

\subsection{Multiplierless relaxation}

Under the constraint of \eqref{beta-cond}, it is possible to force $\beta_i(\r)$ to have values that are signed powers of 2. In this way, all multiplications involved in the product $\bA\bb\up{n}$ of \eqref{sys2b} are reduced to bit shifts. There are various ways to achieve this goal. In this paper, we consider functions $\beta_i(\cdot)$ of the form
\begin{equation}\label{beta-def}
\beta_i(\r):=\rho\big(\lambda\r/T_i\big)
\end{equation}
where $\lambda$ is some chosen constant in $(0,2)$ and
\begin{equation}\label{rho}
\rho(\r):=\sign(\r)\,\max_{2^k\leq|\r|}2^k
\end{equation}
for all $\r\neq0$, with $\rho(0):=0$.
For any $\r>0$, it is clear that $\half\r<\rho(\r)\leq\r$. So $\rho(\r)/\r\in(\half,1]$, and as a result
\begin{equation}\label{msb-cond}
T_i\,\beta_i(\r)/\r=\lambda\,\midfrac{\rho(\lambda\r/T_i)}{\lambda\r/T_i}\in\textstyle(\half \lambda,\lambda]
\end{equation}
for all $\r>0$. By odd symmetry of the function $\rho(\cdot)$, this is also true for all $\r\neq0$. As $\lambda\in(0,2)$, we obtain \eqref{beta-cond} with $\eps=\min(\half\lambda,2{-}\lambda)>0$.

An apparent shortcoming of the function $\beta_i(\cdot)$ of \eqref{beta-def} is that it involves a multiplication and a division. There is a way to avoid them. Note that for any $\r\neq0$ and $\a>0$,
$$\rho(\smallfrac{\r}{\a})=\sign(\r)\,\displaystyle\max_{2^k \a\leq|\r|} 2^k.$$
This value is then found by simple inspection of the binary expansions of $|\r|$ and $\a$. Next, we calculate $\beta_i(\r)$ in the form
\begin{equation}\label{beta-equiv}
\beta_i(\r)=\rho\Big(\midfrac{\r}{T_i/\lambda}\Big).
\end{equation}
The division by $\lambda$ is not eliminated, but $T_i/\lambda$ is to be computed only once for each $i\in\Z$ before the iteration. Moreover, $\lambda$ is only a constant parameter that is roughly and empirically adjusted to accelerate the convergence. We will see in the next section that good results are obtained with a value of $\lambda$ of very low binary complexity.

\subsection{Experimental results}

Under the experimental conditions of Section \ref{sec:exp}, we plot in Fig. \ref{fig2}(e) the performance of the multiplierless relaxation technique we have just devised. At each iteration $n$, the value of  $\|x\up{n}\!-x\|^2$ is reported, where $x\up{n}(t)$ is obtained from \eqref{nth-2} and  $\bc\up{n}$ is recursively obtained from the discrete-time system \eqref{sys2}. In this system, the function $\bB$ is defined by \eqref{B} and \eqref{beta-equiv} where $\lambda$ is taken to be $(2^{-1}{+}2^{-4})^{-1}\simeq 1.8$. In this case, the division $T_i/\lambda$ involved in \eqref{beta-equiv} only requires a few bit shifts and one addition. While the iteration is multiplierless, it yields better results than the full-resolution relaxation-free POCS method of (c), although not as good as the empirically optimized configuration of (d) with constant relaxation.
Meanwhile, it outperforms in error decay rate the result of \cite{Lazar04} in (a).

\section{Real-time circuit implementation}\label{sec:imp}

We saw in Section \ref{sec:mult-free} that the POCS iteration of \eqref{x-rec-rel} is equivalently implemented by iterating the discrete-time system \eqref{sys2} with the multiplierless option of \eqref{B} and \eqref{beta-def}, and injecting the coefficients of the resulting output $\bc\up{n}=(\c_i\up{n})\subi$ into the D/A conversion formula of \eqref{D/A}. In this section, we propose an approximate circuit implementation of the system iteration of \eqref{sys2} as a hardware pipeline of multiplierless time-varying FIR filters. We end the section with experimental results including the effects of FIR windowing and time quantization. We assume from now on that $\Z=\{1,\cdots,N\}$.

\subsection{Approximate iteration}

Concisely, the system \eqref{sys2} amounts to the two-argument transformation
\begin{equation}\label{ideal-iter}
(\br\up{n+1},\bc\up{n+1})=\bR(\br\up{n},\bc\up{n})
\end{equation}
starting from $(\br\up{0},\bc\up{0})=(\bs,\bzero)$, where
$$\bR(\br,\bc):=\big(\br-\bA\bB(\br)\,,\bc+\bB(\br)\big),\qquad\br,\bc\in\RR^N.$$
The transformation $\bB$ defined in \eqref{B} depends on the choice of functions $\beta_1(\r),\cdots,\beta_N(\r)$ and is in general nonlinear. It is however memoryless when thinking of the components of $\br$ as a sequence of time. The issue is the  multiplication by the matrix $\bA$. Although $\bA$ is theoretically of finite size, it is virtually infinite compared to the practical time windows of operation. Now, its coefficients $\langle f_i,f_j\rangle$ typically tend to 0 when $|i{-}j|$ tends to infinity. Like in rectangular windowing for the FIR implementation of lowpass filters, we consider truncating these coefficients as soon as $|i{-}j|$ is larger than some parameter $L\geq0$. This amounts to replacing $\bA$ by the matrix $\hA$ of coefficients
\begin{equation}\label{ski}
\hat\a_{i,j}:=\mbox{\small$\left\{\begin{array}{cl}
\langle f_i,f_j\rangle,&i,j\in\Z~\mbox{and}~|i{-}j|\leq L\\0,&\mbox{otherwise}
\end{array}\right.$}.
\end{equation}
So, in real implementation, \eqref{ideal-iter} is replaced by
\begin{equation}\label{approx-iter}
(\br\up{n+1},\bc\up{n+1})=\hR(\br\up{n},\bc\up{n})
\end{equation}
where
\begin{equation}\label{approx-map}
\hR(\br,\bc):=\big(\br-\hA\bB(\br)\,,\bc+\bB(\br)\big),\qquad\br,\bc\in\RR^N.
\end{equation}

\subsection{Sliding-window pipeline implementation}\label{pipe1}

\begin{figure}
\centerline{\scalebox{0.7}{\includegraphics{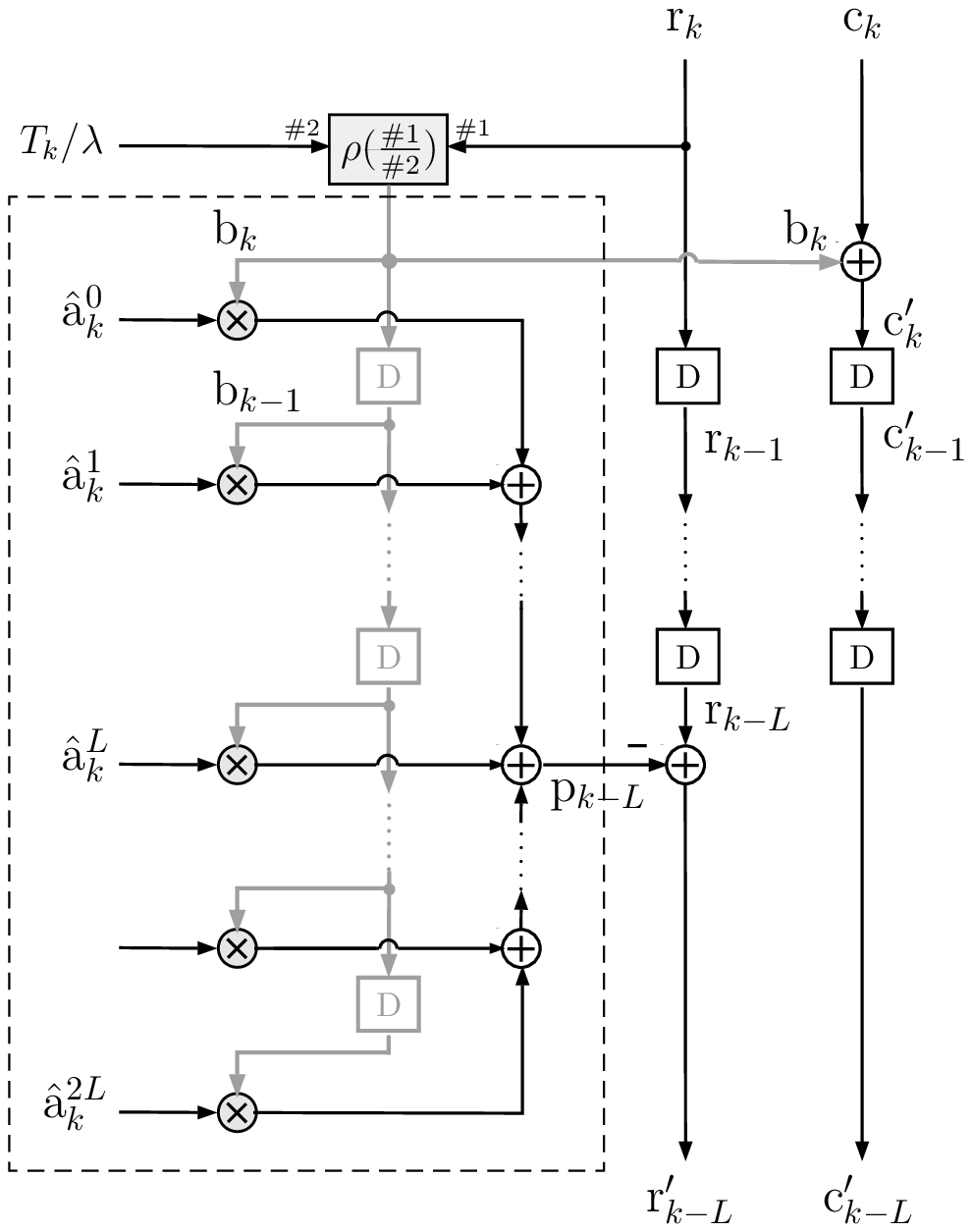}}}
\centerline{(a)}
\vspace{4mm}
\centerline{\scalebox{0.85}{\includegraphics{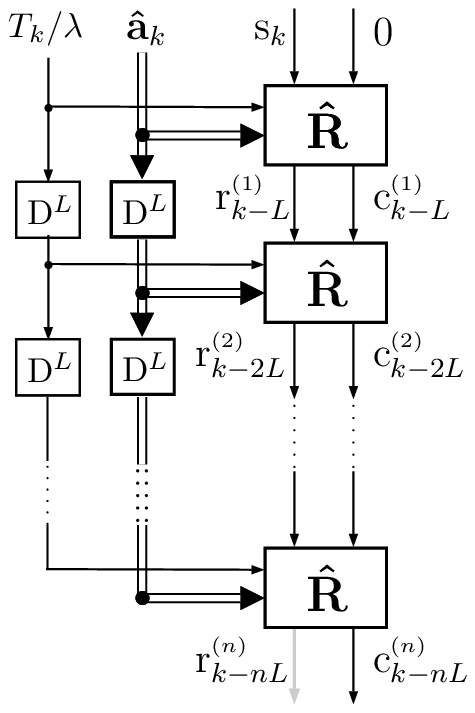}}}
\centerline{(b)}
\vspace{-1mm}
\caption{Pipeline implementations: (a) Operation $(\br',\bc')=\hR(\br,\bc)$ with the functions $\beta_i$ of \eqref{beta-def}; (b) Operation  $(\br\up{n},\bc\up{n})=\hR^n(\bs,\bzero)$. The inputs $\hat\a_k^\ell$ are defined in \eqref{sjn} and
$\hba_k:=\big(\hat\a^0_k,\hat\a^1_k,\cdots,\hat\a^{2L}_k\big)$. The gray lines in (a) highlight the connections conveying signals in signed power-of-2 format, and the operators shaded in gray imply a multiplication or division by a signed power of 2. \label{fig:pipe}}
\end{figure}

We show in Fig. \ref{fig:pipe}(a) a real-time pipeline implementation of the single transformation $(\br',\bc')= \hR(\br,\bc)$. It is derived as follows. From \eqref{approx-map}, we have
\begin{eqnarray}
\br'=\br-\bp\quad&\mbox{and}&\quad\bc'=\bc+\bb\label{rp}\\
\mbox{where}\qquad\bp:=\hA\bb\quad&\mbox{and}&\quad\bb:=\bB(\br).\qquad\qquad\nonumber
\end{eqnarray}
Using explicitly the multiplierless functions $\beta_i$ of \eqref{beta-equiv}, the components of $\bb$ are
\begin{equation}\label{b-def}
\b_i=\beta_i(\r_i)=\rho\Big(\midfrac{\r_i}{T_i/\lambda}\Big).
\end{equation}
Meanwhile, the components of $\bp$ are
\begin{equation}\nonumber
\p_k=\textstyle\sum\limits_{j=k-L}^{k+L}\hat\a_{k,j}\,\b_j.
\end{equation}
Note that $\p_k$ depends on $\b_{k+L}$. So at a given instant $k$, only $\p_{k-L}$ can be obtained in a causal manner.
We have
\begin{equation}\label{causal-trunc}
\p_{k-L}=\textstyle\sum\limits_{j=k-2L}^k\hat\a_{k-L,j}\,\b_j=
\sum\limits_{\ell=0}^{2L}\hat\a_k^\ell\,\b_{k-\ell}
\end{equation}
where for each $\ell\in\{0,{\cdots},2L\}$,
\begin{align}\label{sjn}
\hat\a_k^\ell&:=\hat\a_{k-L,k-\ell}=
\mbox{\small$\left\{\begin{array}{cl}
\!\!\langle f_{k-L},f_{k-\ell}\rangle,& k{-}L,k{-}\ell\in\Z\\
0,&\mbox{otherwise}
\end{array}\right.$}.
\end{align}
Equations \eqref{rp}, \eqref{b-def} and \eqref{causal-trunc} can then be mapped to the block diagram of Fig. \ref{fig:pipe}(a). Each node signal is a function of the discrete-time index $k$, which is incremented in real time from $k{-}1$ at the switching instant $t_k$. The symbol $\mathrm{D}$ represents the delay operation with respect to $k$. The dashed frame highlights the structure of time-varying FIR filter operating on the sequence $(\b_k)_{k\in\Z}$ of signed powers of 2.

Fig. \ref{fig:pipe}(b) shows the global pipeline architecture for the computation of $(\br\up{n},\bc\up{n})=\hR^n(\bs,\bzero)$. The operation $\mathrm{D}^L$ is the delay by $L$ discrete-time instants. We will show in Section \ref{sec:lookup-table} how the coefficients $\hat\a_k^\ell$ can be obtained in real time by table lookup.

\subsection{Relaxed bandlimitation}\label{sec:rel-band}

The coefficients $\langle f_i,f_j\rangle$ are expected to decay with
\begin{equation}\label{Tij}
T_{i,j}:=t_i-t_j
\end{equation}
at the slow rate of $1/|T_{i,j}|$ due to their connection to the sinc function. As a classically known phenomenon, a plain truncation of such a sequence of coefficients is expected to induce disappointingly large errors. Advanced techniques of windowing are available for linear and time-invariant DSP, but not for the present case of time-varying operations. Moreover, the truncated operator $\hR$ is iterated, making the process sensitive to in-band distortions. With the lack of knowledge in this problem, we propose to maintain the abrupt truncation of the coefficients $\langle f_i,f_j\rangle$ but relax the bandlimitation of $f_i(t)$ by taking instead of \eqref{fi},
\begin{equation}\label{relax-band}
f_i(t)=\varphi(t)*\pi_i(t)
\end{equation}
where $\varphi(t)$ is the impulse response of a non-ideal lowpass filter with faster decay than the sinc function. Specifically, we maintain the flat in-band frequency response of $\varphi(t)$ but allow a smooth cutoff transition (of cosine type) between the angular frequencies of $\pi$ and $r\pi$ for some coefficient $r>1$. With a faster decay rate, the purpose is to limit the damages due to truncation and eventually limit in-band distortions. Mathematically, this amounts to replacing $P_\scB$ in \eqref{x-rec1} by a non-ideal bandlimitation. One will naturally expect degradations in the efficiency of the POCS's.

\subsection{Experimental results}\label{sec:TEM2}

\begin{figure}
\includegraphics[width=0.99\columnwidth]{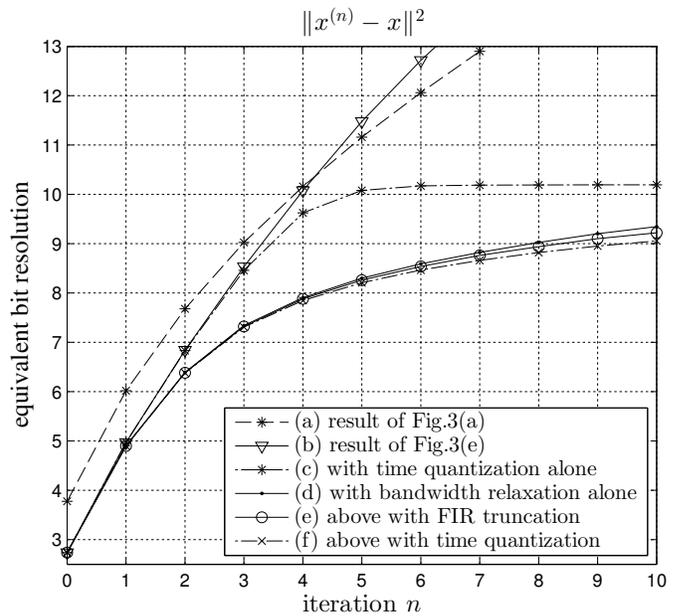}
\caption{In-band MSE of $n$th reconstruction estimate $x\up{n}$ under the experimental conditions of Fig. \ref{fig2} with additional non-idealities: (a) method of \cite{Lazar04} (reproduced from Fig. \ref{fig2}(a)); (b) ideal multiplierless POCS (reproduced from Fig. \ref{fig2}(e)); (c) with time quantization (step size = $2^{-12}$); (d) with relaxed bandlimitation ($r=1.4$); (e) with relaxed bandlimitation ($r=1.4$) and FIR truncation ($L=17$); (f) with relaxed bandlimitation ($r=1.4$), FIR truncation ($L=17$) and time quantization (step size = $2^{-12}$).\label{fig3}}
\end{figure}

We show in Fig. \ref{fig3} the effect of the various practical approximations on the multiplierless reconstruction scheme of Fig. \ref{fig2}(e), which is reproduced as curve (b) in Fig. \ref{fig3}. For reference, we have also reported in curve (a) the result of Fig. \ref{fig2}(a) obtained from the method of \cite{Lazar04}. We report in (d) the performance degradation due to bandwidth relaxation alone with $r=1.4$, as presented in the previous section. Under this condition, we next apply the truncation approximation of \eqref{approx-iter} with $L=17$, which yields the result of curve (e). Although the experiment is performed on an input of period 257, it is representative of aperiodic inputs as the window of operation resulting from the truncation is only of approximate length 19 in average, which is small compared to the input period. As shown in the figure, 6 iterations are needed to obtain a reconstruction resolution of 8.5 bits. The total number of adders required by the system for $n$ iterations is $n(2L{+}2)+(6L{+}3)$ where $2L{+}2$ is the complexity of $\hR$ in Fig. \ref{fig:pipe}(a) and $6L{+}3$ is the required complexity to compute the multidimensional input
\begin{equation}\label{hba}
\hba_k:=\big(\hat\a^0_k,\hat\a^1_k,\cdots,\hat\a^{2L}_k\big)
\end{equation}
as will be shown in Section \ref{sec:lookup-table}. With $L=17$ and $n=6$, this implies 322 adders. Roughly, we have observed that each additional bit of reconstruction resolution requires a doubling of the computation complexity. According to our observations, the bottleneck of reconstruction accuracy is the slow decay of the sinc function required for exact bandlimitation.

It is also interesting to see the behavior of the algorithm with additional noise. As a concrete source of noise, we choose the quantization in time of the switching instants $\tau_n$ of the encoder. This implies errors on both $t_i$ and $\s_i$ as can be seen in \eqref{tn-xn}. We show the resulting additional degradation in curve (f) with the time-quantization step size of $2^{-12}$. This time resolution has been chosen by observing its effect in absence of all other distortions, as shown in curve (c). In fact, time quantization is necessary not only for digital processing, but also to limit the possible values of $\langle f_i,f_j\rangle$ to a finite number so that they can be precalculated and stored in a lookup table. According to a method presented in Section \ref{sec:lookup-table} and the signal statistics of the present experiment, this lookup table is evaluated to fit in a memory of less than 100 KB.

Overall, this experiment is an initial demonstration of the effects of practical non-idealities on the POCS algorithm, including FIR truncation, bandwidth relaxation and time quantization. The performance degradations compared to the ideal algorithm appear to be mostly from the truncation of sinc-like functions, which is an unavoidable obstacle when dealing with the finite-complexity processing of bandlimited functions. The new difficulty is time-varying filter windowing for which little knowledge is available. The results presented here are mostly preliminary, with potential improvements from future investigations on time-varying filtering.

\section{Matrix coefficients by table lookup}\label{sec:lookup-table}

Until now, we have assumed the inner-products $\langle f_i,f_j\rangle$ to be available. In the previous section, they are involved in \eqref{sjn}. In steady state and more precisely for all $k=2L{+}1,\cdots,N$, we simply have
\begin{equation}\label{steady-state-akl}
\hat\a_k^\ell=\langle f_{k-L},f_{k-\ell}\rangle,\qquad\ell=0,\cdots,2L.
\end{equation}
Based on an idea introduced in \cite{Hand12} and following more elaborate derivations from \cite{chap3}, we show that the coefficients $\langle f_i,f_j\rangle$ can be obtained from a single analytical function $h(t)$ applied to time distances $T_{i,j}=t_i-t_j$ as defined in \eqref{Tij}. With time quantization, the values of this function can be stored in a lookup table. We also propose a pipeline circuit implementation to obtain in real time the required differences $T_{i,j}$ from the sampling-step sequence $(T_i)\subi$ of \eqref{Tm}.

\subsection{Expression of $\langle f_i,f_j\rangle$}\label{pure-int}

We consider the more general expression of $f_i(t)$ from \eqref{relax-band} to allow the use of functions of faster decay as was motivated in Section \ref{sec:rel-band}. Let us define
\begin{equation}\label{phi2}
a_\varphi(t):=\varphi(t)*\varphi({-}t).
\end{equation}

\begin{proposition}
\begin{equation}\label{sksi-int}
\langle f_i,f_j\rangle=h( T_{i,j-1})-h( T_{i-1,j-1})-h( T_{i,j})+h( T_{i-1,j})
\end{equation}
where $T_{i,j}:=t_i-t_j$ as defined in \eqref{Tij} and
\begin{equation}\label{h}
h(t)=\int_0^t(t{-}\tau)\, a_\varphi(\tau)\,\dif\tau.
\end{equation}
\end{proposition}
\medskip
\begin{IEEEproof}
We have
$\textstyle\langle f_i, f_j\rangle=\big\langle\varphi\,{*}\,\pi_i,\varphi\,{*}\,\pi_j\big\rangle=
\big\langle \pi_i, a_\varphi\,{*}\,\pi_j\big\rangle
=\int_{t_{i-1}}^{t_i}( a_\varphi\,{*}\,\pi_j)(t)\dif t$.
Next,
$\textstyle( a_\varphi\,{*}\,\pi_j)(t)=\int_{t_{j-1}}^{t_j} a_\varphi(t{-}\tau)\dif\tau
=\psi(t{-}t_{j-1})-\psi(t{-}t_j)$
where $\psi(\tau):=\int_0^\tau a_\varphi(s)\,\dif s$. Thus,
$$\langle f_i, f_j\rangle=
\smallint{t_{i-1}}{t_i}\psi(t{-}t_{j-1})\dif t-\smallint{t_{i-1}}{t_i}\psi(t{-}t_j)\dif t.$$
Defining  $h(t):=\int_0^t \psi(\tau)\dif\tau$, we have for any $k$,
$\int_{t_{i-1}}^{t_i}\psi(t{-}t_k)\dif t=h(t_i{-}t_k)-h(t_{i-1}{-}t_k)=h( T_{i,k})-h( T_{i-1,k})$. This leads to \eqref{sksi-int}. Since $h(t)=\int_0^t\int_0^\tau a_\varphi(s)\,\dif s\,\dif\tau$, one obtains \eqref{h} from the Cauchy formula for the second repeated integral of $a_\varphi(t)$ (derived by integration by part noting that $a_\varphi(\tau)=\psi'(\tau)$).
\end{IEEEproof}
\medskip
A slight numerical issue with \eqref{sksi-int} is that $\lim_{|t|\rightarrow\infty}h(t)=\infty$, while $\lim_{|i-j|\rightarrow\infty}\langle f_i,f_j\rangle=0$. We show in Appendix \ref{num-trick} how this problem can be circumvented.

\subsection{Real-time computation of $\hba_k$}

To obtain the coefficients $\hat\a_k^\ell$, we need to express $\langle f_{k-L},f_{k-\ell}\rangle$ as required by \eqref{sjn}. Let us define the coefficients
\begin{equation}\label{fkl}
\h_k^\ell:=h(T_{k-L,k-\ell}).
\end{equation}
After verifying that $h(T_{k-L-i',k-\ell-j'})=\h_{k-i'}^{\ell-i'+j'}$ and taking various values of $i',j'\in\{0,1\}$, one easily obtains from \eqref{steady-state-akl} and \eqref{sksi-int} that
\begin{equation}\label{sjn2}
\hat\a_k^\ell=\h^{\ell+1}_k-\h^{\ell}_{k{-}1}-\h^{\ell}_{k}+\h^{\ell-1}_{k{-}1}.
\end{equation}
The values of $\h_k^\ell$ in \eqref{fkl} can be obtained from the time values $T_{k-L,k-\ell}$ by table lookup.

\begin{figure}
\centerline{\scalebox{0.84}{\includegraphics{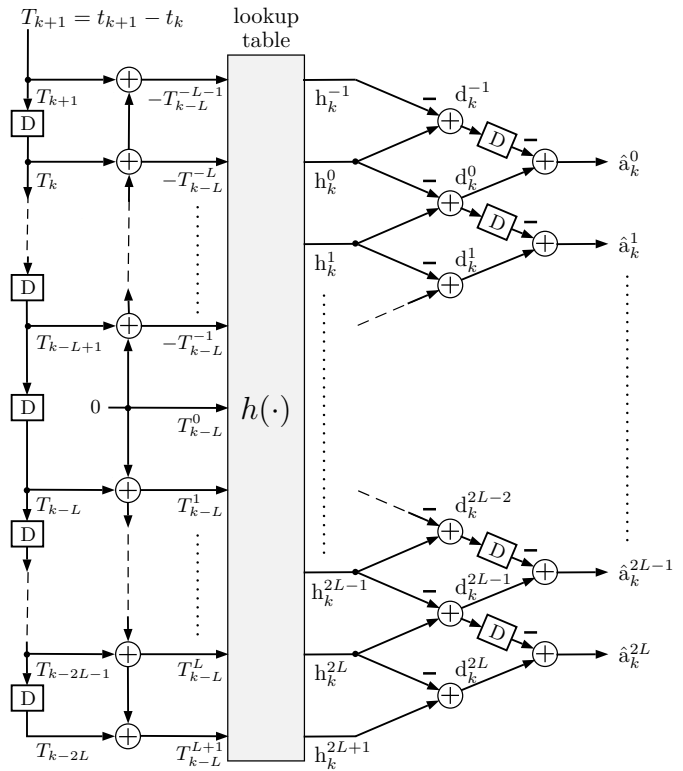}}}
\caption{Synchronous use of lookup table to obtain $\hba_k=\big(\hat\a^0_k,\hat\a^1_k,\cdots,\hat\a^{2L}_k\big)$ from $T_{k+1}=t_{k+1}{-}t_k$. The inputs to the table are obtained from \eqref{id12}, and the postprocessing of the table outputs results from \eqref{table-output}.\label{fig:table-read}}
\end{figure}
A difficulty is the real-time transformation of the sequence of switching instants $(t_i)_{0\leq i\leq N}$ into the required values $T_{k-L,k-\ell}$.
In practice, the time encoder typically provides this sequence in the form of the successive differences\footnote
{As a basic practical technique, the time quantized value of $T_k$ is provided by a counter that is incremented at a fast clock rate and is reset to 0 right after each instant $t_k$. The time quantization step size is defined by the clock period.} $T_k=t_k-t_{k-1}$ defined in \eqref{Tm}.
We show in Fig. \ref{fig:table-read} how the sequence $T_{k+1}$ can be manipulated in real discrete time to eventually output the required values of $T_{i,j}$ for $\hba_k$. The proposed technique is to consider the generalized sequence
\begin{equation}\label{Dkn}
 T^n_k:=T_{k,k-n}=t_k-t_{k-n}
\end{equation}
and use the relations
\begin{equation}\label{id12}
\textstyle\sum_{j=\ell}^{n-1}T_{k-j}=T_{k-\ell}^{n-\ell}=-T_{k-n}^{\ell-n}
\end{equation}
easy to verify from \eqref{Dkn}. With \eqref{sjn2}, \eqref{fkl} and the even symmetry of $h(\cdot)$ easy to check, $\hat\a_k^\ell$ can then be obtained from the lookup table by the successive operations
\begin{equation}\label{table-output}
\h_k^\ell=h\big(\pm T_{k-L}^{\ell-L}\big),\quad\d_k^\ell=\h_k^{\ell+1}-\h_k^{\ell},\quad \hat\a_k^\ell=\d_k^\ell-\d_{k-1}^{\ell-1}.
\end{equation}
The global system requires $6L{+}3$ adders.

Assuming that the sequence $T_k$ is bounded, the argument $T_{k-L,k-\ell}$ to the
function $h(\cdot)$ of \eqref{fkl} remains bounded. With time quantization, it can
therefore only take a finite number of values, thus allowing a
lookup table of finite size. In the experiment of Section \ref{sec:TEM2}, we recall that its size was evaluated to be less than 100 KB.

\section{Preliminary analysis of data noise effect}\label{sec:noise}

We had a glimpse at the behavior of the algorithm with some data noise in the experiment of Section \ref{sec:TEM2} and more specifically in Fig. \ref{fig3}(c). It would be desirable to get a little more analytical insight on the effect of noise on the estimates, especially given the tendency for sampling to generate ill-conditioned operators \cite{Choi98}. Given the difficulty of the analysis, we will limit ourselves to POCS reconstruction without relaxation. We keep the assumption that $\Z$ is finite.

\subsection{Orthogonal error decomposition}

When one only has access to the noise corrupted sampling sequence $\hbs=\bs+\bn$ of \eqref{noise}, we mentioned in Section \ref{sec:samp-op} that one is only left with the noisy POCS iteration
$$x\up{n+1}=R_\hbs\,x\up{n}.$$
We saw in \eqref{noise-lim} that these iterates
tend to the deviated reconstruction $x_\bbs(t)$. We wish to have some insight on the iterated error signal
$$e\up{n}:=x\up{n}-x_\bs$$
in terms of $\bn$. It is easy to see from \eqref{op-expr} and \eqref{noise} that
$$R_\hbs u=R_\bs u+S^*\bn$$
for all $u\in\scB$. Then,
$$x\up{n+1}-x_\bs=R_\bs x\up{n}+S^*\bn-R_\bs x_\bs$$
since $x_\bs(t)$ is a fixed point of $R_\bs$. With \eqref{linear-part}, we obtain
\begin{equation}\label{e-fix}
e\up{n+1}=Me\up{n}+S^*\bn.
\end{equation}
One can see from \eqref{op-expr} that $M$ is a self-adjoint operator on $\scB$. As the particular case of $M^\lambda$ with $\lambda=1$, we know from Section \ref{sec:frame} that $M$ leaves $\scV_f$ invariant and from Theorem \ref{theo:ineq} that $\|M u\|<\|u\|$ for all $u\in\scV_f\backslash\{0\}$. We conclude that $\scV_f$ yields an orthonormal basis $(\psi_i)_{i\in\Z}$ of eigenvectors of $M$ of real eigenvalues $(\mu_i)_{i\in\Z}$ such that
$$|\mu_i|<1,\qquad\forall i\in\Z.$$
Defining the components of $e\up{n}$ and $S^*\bn$ in this basis
$$\e_i\up{n}:=\langle\psi_i,e\up{n}\rangle\qquad\mbox{and}\qquad
\n_i:=\langle\psi_i,S^*\bn\rangle,\qquad\forall i\in\Z$$
then \eqref{e-fix} implies that
\begin{equation}\label{err-iter}
\e_i\up{n+1}=\mu_i\e_i\up{n}+\n_i,\qquad\forall i\in\Z.
\end{equation}

\subsection{Semi-convergence analysis}

Since $x\up{0}=0$, the initial error is
$$e\up{0}=-x_\bs$$
and is solely dependent on the ideal reconstruction target. Meanwhile, the final error signal is
$$e\up{\infty}=x_\bbs-x_\bs$$
as a result of \eqref{noise-lim}, and gives the pure deviation of the algorithm from noise. Since $\e_i\up{\infty}=\mu_i\e_i\up{\infty}+\n_i$ from \eqref{err-iter}, the $i$th component of $e\up{\infty}$ is then
$$\e_i\up{\infty}=\frac{\n_i}{1-\mu_i},\qquad i\in\Z.$$
The noise component $\n_i$ is attenuated when $\mu_i<0$, but gets particularly amplified when $\mu_i$ is close to 1, which happens when the sampling is badly conditioned. In the experimental condition of Fig. \ref{fig2}(c) however, we find numerically that ${|\mu_i|<0.3}$ for all $i\in\Z$, among which less than 1\% satisfy ${\mu_i>0.17}$. This shows the good conditioning of the time encoding machine, and hence implies its good behavior with respect to noise.

But it is interesting to see in more details how $\e_i\up{n}$ moves between $\e_i\up{0}$ and $\e_i\up{\infty}$. Since $\e_i\up{\infty}$ is a fixed point of \eqref{err-iter}, then $\e_i\up{n+1}\!-\e_i\up{\infty}=\mu_i(\e_i\up{n}\!-\e_i\up{\infty})$. By induction, one finally finds that
\begin{equation}\label{err-lim}
\e_i\up{n}=\mu_i^n\,\e_i\up{0}+(1-\mu_i^n)\,\e_i\up{\infty},\qquad i\in\Z.
\end{equation}
The first term gives the zero-noise component of the error and corresponds to the intrinsic convergence behavior of the algorithm. The second term isolates the contribution of data noise in the iteration. This type of error decomposition is typically performed in the semi-convergence analysis of an algorithm \cite{Elfving14}. As $|\mu_i|<1$, one sees the exact analytical law under which $\e_i\up{n}$ moves from $\e_i\up{0}$ to $\e_i\up{\infty}$. We saw that $\e_i\up{\infty}$ may be undesirably amplified when $\mu_i$ is close to 1. But in this case, one notices that more iterations are needed for the noise term in \eqref{err-lim} to reach its full value. The action of stopping the iteration at an early enough stage thus plays a role of reconstruction regularization \cite{Landi16} under critically ill-conditioned sampling.

\section{Summary and discussion}

The contribution of this paper is two-fold. The first part is theoretical. Based on POCS, we proposed an algorithm that systematically converges to the unique minimal-norm bandlimited signal yielding a given ASDM output, with absolutely no assumption on the sampling condition. In the presence of noise, the reconstruction coincides with the pseudo-inversion of the linear operator induced by the time encoding. The second part is practical. While the typical approach to signal reconstruction from non-uniform samples is to perform block-based ill-conditioned algebraic inversions, our method returns to the more traditional signal method of sliding-window processing, in the form of time-varying multiplerless FIR filters. While avoiding ill-posed algebra, our algorithm however has to face the traditional difficulty of filter windowing which is likely to play a major part in the bottleneck of performance.
This comes with the new issues of sliding-window truncation under time-varying signal processing, and filter non-idealities in the iterative process of POCS. At this stage, the practical numerical results of this paper are only preliminary, with future potential for improvements after further theoretical investigations of these non-trivial problems.

\appendix

\subsection{2-periodically nonuniform sampling}\label{app:2per}

We consider in this appendix the case where $t_i$ is of the form of \eqref{2per-ti} for some constant $\delta\in[0,\half)$.
\pp

\subsubsection{Input example}

For illustration, we first give an example of ASDM input that yields the switching instants of \eqref{2per-ti} when $d\in(0,\frac{1}{4})$. Under the condition that $d=f(\delta):=(1-2\delta\sin(\delta\pi))/4$ for some $\delta\in(0,\half)$, it can be verified that the input $x(t):=\delta\pi\cos(\pi t)$ satisfies \eqref{int-equ} with $\tau_{2i}=i+(-1)^i\delta$ and $\tau_{2i+1}=i+\half$ for all $i\in\ZZ$. Since $t_i=\tau_{2i}$, then \eqref{2per-ti} is satisfied for all $i$. Now, in practice, $d$ has a fixed value imposed by the circuit. It can be seen that $f(\delta)$ maps $(0,\half)$ into $(0,\frac{1}{4})$ in a strictly decreasing manner. So when $d\in(0,\frac{1}{4})$, there exists a unique $\delta\in(0,\half)$ such that $d=f(\delta)$. Then, \eqref{2per-ti} is achieved with this value of $\delta$ and $x(t):=\delta\pi\cos(\pi t)$.

\pp
\subsubsection{Condition of perfect reconstruction}\label{app:2per-rec}

By Fourier analysis, we are going to show that $(\s_i)_{i\in\ZZ}$ in \eqref{xi} uniquely characterizes the bandlimited input $x(t)$. Because the nonuniformity of $(t_i)_{i\in\ZZ}$ is 2-periodic, we have $f_{2k+i}(t)=f_i(t{-}2k)$ for any $i,k\in\ZZ$. Splitting $(\s_i)_{i\in\ZZ}$ into the two sequences
\begin{equation}\label{y01}
y_0(k):=\s_{2k}\qquad\mbox{and}\qquad y_1(k):=\s_{2k+1},\qquad k\in\ZZ,
\end{equation}
we then obtain
$$y_i(k)=\big\langle f_i(t{-}2k),x(t)\big\rangle,\qquad k\in\ZZ,~i\in\{0,1\}.$$
Now, since $(\sinc(t{-}n))_{n\in\ZZ}$ is an orthonormal basis of $\scB$, we have for any $u(t),v(t)\in\scB$ the inner-product preservation
$$\langle u,v\rangle=\big\langle u(n),v(n)\big\rangle_{\ell^2}:=\smallsum{n\in\ZZ}u(n)v(n).$$
$$\mbox{So,}\quad y_i(k)=\big\langle f_i(n{-}2k),x(n)\big\rangle_{\ell^2},\quad k\in\ZZ,~i\in\{0,1\}.\qquad$$
These are the equations of the analytical section of a 2-channel filter bank \cite[\S3.2.1]{Vetterli95}. Let the discrete-time Fourier transform $U(\om)$ of a sequence $u(n)$ be defined as
$$U(\om):=\smallsum{n\in\ZZ}u(n)\,e^{-j\om n},\qquad\forall\om\in[-\pi,\pi].$$
If $(u(n))_{n\in\ZZ}$ are the Nyquist samples of $u(t)\in\scB$, note that $U(\om)$ is also the continuous-time Fourier transform of $u(t)$ within the baseband $[-\pi,\pi]$. One obtains from \cite{Vetterli95} the relation\footnote
{This relation is given in \cite[\S3.2.1]{Vetterli95} in the $z$-domain with $h_i[n]=f_i(-n)$.}
\begin{eqnarray}
&\mbox{\small$\begin{bmatrix}
Y_0(2\om)\\
Y_1(2\om)
\end{bmatrix}$}
=\midhalf\,\bF(\om)^*\,
\mbox{\small$\begin{bmatrix}
X(\om)\\
X(\om{-}\pi)
\end{bmatrix}$},\qquad\forall\om\in[0,\pi]\label{XY}\\
\lefteqn{\mbox{where}}\hspace{5mm}&
\bF(\om):=\mbox{\small$\begin{bmatrix}
F_0(\om)&F_1(\om)\\
F_0(\om{-}\pi)&F_1(\om{-}\pi)
\end{bmatrix}$}.&\hspace{5mm}\nonumber
\end{eqnarray}
Because rectangular functions have continuous Fourier transforms, $\bF(\om)$ is continuous in $[0,\pi]$. It is then sufficient that $\bF(\om)$ be invertible for each $\om\in[0,\pi]$ for $X(\om)$ to be stably recoverable in $[-\pi,\pi]$ from \eqref{XY}.
We have
\begin{align*}
f_0(t)&=\sinc(t)*1_{I_\delta}(t+\smallhalf)\\
f_1(t)&=\sinc(t)*1_{I_{-\delta}}(t-\smallhalf)
\end{align*}
where $I_\alpha:=[-\half{-}\alpha,\half{+}\alpha)$ for any $\alpha\in(-\half,\half)$.
Then
$$F_i(\om)=e^{j(-1)^i\om/2}\,\midfrac{\sin\left(T_i\,\om/2\right)}{\om/2},
\quad\forall\om\in[-\pi,\pi],\,i=0,1$$
where $T_i=1+(-1)^i2\delta$.
By symbolic computation software, we obtain
$$\det(\bF(\om))=4j\cos(\delta\pi)\frac{\sin(\om)}{\om(\pi-\om)},\qquad\forall\om\in[0,\pi].$$
This is never 0 since $\delta\in[0,\half)$ and $\sin(\om)/\om/(\pi{-}\om)\geq 4/\pi^2$ for all $\om\in[0,\pi]$. Thus $x(n)$ can be retrieved from the sequences $y_0(k)$ and $y_1(k)$, and hence $x(t)$ can be uniquely recovered from $(\s_i)_{i\in\ZZ}$.
\pp

\subsubsection{Contracting algorithm}\label{app:2per-contract}

The goal is to analyze the norm of the mapping $M$ of \eqref{Mu} with the functions $g_i(t)$ of \eqref{gi} adopted in \cite{Lazar04}. We start by analyzing
\begin{equation}\label{xhx}
\hat x(t):=\smallsum{i\in\ZZ}\langle\pi_i,x\rangle\,g_i(t).
\end{equation}
Similarly to $f_i(t)$, we have $g_{2k+i}(t)=g_i(t{-}2k)$ for any $i,k\in\ZZ$. From \eqref{xn2} and \eqref{y01}, we then obtain
$$\hat x(t)=\smallsum{k\in\ZZ}y_0(k)\,g_0(t-2k)+\smallsum{k\in\ZZ}y_1(k)\,g_1(t-2k).$$
Restricting $t$ to the Nyquist sampling instants $n\in\ZZ$, this is the equation of the synthesis section of a 2-channel filter bank. One can derive from \cite[\S3.2.1]{Vetterli95} that
\begin{equation}\nonumber
\hat X(\om)=
\mbox{\small$\begin{bmatrix}
G_0(\om)&G_1(\om)\end{bmatrix}$}\,
\mbox{\small$\begin{bmatrix}
Y_0(2\om)\\
Y_1(2\om)
\end{bmatrix}$},\qquad\forall\om\in[-\pi,\pi].
\end{equation}
As $Y_i(2\om)$ is $\pi$-periodic, then
\begin{eqnarray}
&\mbox{\small$\begin{bmatrix}
\hat X(\om)\\
\hat X(\om{-}\pi)
\end{bmatrix}$}
=\bG(\om)\,
\mbox{\small$\begin{bmatrix}
Y_0(2\om)\\
Y_1(2\om)
\end{bmatrix}$},\qquad\forall\om\in[0,\pi]\label{YhX}\\
\lefteqn{\mbox{where}}\hspace{8mm}&
\bG(\om):=\mbox{\small$\begin{bmatrix}
G_0(\om)&G_1(\om)\\
G_0(\om{-}\pi)&G_1(\om{-}\pi)
\end{bmatrix}$}.&\hspace{7mm}\nonumber
\end{eqnarray}
Now, let
$$z(t):=Mx(t)=x(t)-\hat x(t)$$
according to \eqref{Mu} and \eqref{xhx}.
By combining \eqref{XY} and \eqref{YhX}, we obtain
\begin{eqnarray*}
&\mbox{\small$\begin{bmatrix}
Z(\om)\\
Z(\om{-}\pi)
\end{bmatrix}$}
=\bM(\om)\,\mbox{\small$\begin{bmatrix}
X(\om)\\
X(\om{-}\pi)
\end{bmatrix}$},\qquad\forall\om\in[0,\pi]\\
\lefteqn{\mbox{where}}\hspace{7mm}&
\bM(\om):=\bI-\smallhalf\,\bG(\om)\,\bF(\om)^*&\hspace{7mm}
\end{eqnarray*}
and $\bI$ is the identity matrix of size 2. It can be shown that
$$\|M\|=\max_{\om\in[0,\pi]}\|\bM(\om)\|_2$$
where $\|\cdot\|_2$ is here the matrix induced 2-norm.

We now consider explicitly the functions $g_i(t)$ of \eqref{gi}.
We simply have $G_i(\om)=e^{-j\om \bar t_i}$ with $\bar t_i=(-1)^{i-1}\half$ for $i=0,1$. We find numerically that $\|\bM(\om)\|_2$ is maximized at $\om=0$. By symbolic computation software, we obtain that
$$\|\bM(0)\|_2^2=h(\delta):=4\delta^2+\big(1{-}\smallfrac{2}{\pi}\cos(\delta\pi)\big)^2.$$
As $\frac{2}{\pi}<1$, it is easy to see that $h(\delta)$ is the sum of two increasing functions of $\delta$ in $[0,\half]$, with $h(0)=(1{-}\smallfrac{2}{\pi})^2<1$ and $h(\half)=2$. It is found numerically that $h(\delta_0)=1$ for $\delta_0=0.351..$. This implies that $\|M\|>1$ when $\delta>0.352$, and hence when $T_\m=1+2\delta>1.72$.

\subsection{Proof of Proposition \ref{prop:x*charac}}\label{app:x*charac}

Let $\bar x$ be the orthogonal projection of $x$ onto $\scV_f$. Since $\bar x-x\in\scV_f^\perp$ by construction, \eqref{SV} implies $\bar x\in\scS_\bs$. For any $u\in\scS_\bs$, $u-\bar x=(u-x)+(x-\bar x)\in\scV_f^\perp$ due to \eqref{SV} again. So $u-\bar x$ is orthogonal to $\bar x-v$ for any $v\in\scV_f$. By the Pythagorian theorem, we conclude that
\begin{equation}\label{Pyth}
\forall u\in\scS_\bs,v\in\scV_f,\quad \|u-v\|^2=\|u-\bar x\|^2+\|\bar x-v\|^2.
\end{equation}
With $u=x_\bs$ and $v=0$, we obtain $\|x_\bs\|^2=\|x_\bs{-}\bar x\|^2+\|\bar x\|^2$, which implies $\bar x=x_\bs$ since $\bar x\in\scS_\bs$ and due to \eqref{x*def}. For $u\in\scS_\bs$ given, \eqref{Pyth} also shows that $\|u-v\|$ is minimized with $v$ when $v=\bar x$, which proves (i). Finally, if we take $u\in\scS_\bs\cap\scV_f$, then \eqref{Pyth} with $v=u$ implies that $u=\bar x$, which proves (ii).

\subsection{Proof of Theorem \ref{theo:ineq}}\label{app:ineq}

From \eqref{Mu2}, \eqref{our-gi} and \eqref{fi}, $M^\blambda u=P_\scB Q^\blambda u$
for all $u\in\scB$,  where
$$Q^\blambda u:=u-\textstyle\sum\limits_{i\in\Z}\lambda_i\langle\pi_i,u\rangle\pi_i/\|\pi_i\|^2=u-\sum\limits_{i\in\Z}\lambda_i\langle \hat \pi_i,u\rangle\,\hat \pi_i$$
and $\hat \pi_i:=\pi_i/\|\pi_i\|$. So $\|M^\blambda u\|\leq\|Q^\blambda u\|$. Let $\scV_\pi$ be the closed linear span of $(\pi_i)_{i\in\Z}$. It is easy to see that $Q^\blambda$ leaves $\scV_\pi$ invariant and is identity in $\scV_\pi^\perp$.
Let $u\in\scV_f\backslash\{0\}$. Writing the decomposition of $u=v+w$ in $\scV_\pi\oplus\scV_\pi^\perp$, one obtains
\begin{equation}\label{double-Pyth}
\|u\|^2=\|v\|^2+\|w\|^2~~\mbox{and}~~
\|Q^\blambda u\|^2=\|Q^\blambda v\|^2+\|w\|^2.
\end{equation}
Since $u\in\scV_f\backslash\{0\}$, $\langle f_{i_0},u\rangle\neq0$ for some $i_0\in\Z$. Due to \eqref{inner-id}, $\langle \pi_{i_0},v\rangle=\langle \pi_{i_0},u\rangle=\langle f_{i_0},u\rangle\neq0$. As $v$ yields the expansion $v=\sum_{i\in\Z}\langle \hat \pi_i,v\rangle\hat \pi_i$ by orthonormality of $(\hat \pi_i)_{i\in\Z}$, then
$Q^\blambda v=\sum_{i\in\Z}(1{-}\lambda_i)\langle \hat \pi_i,v\rangle\hat \pi_i$. Thus
\begin{equation}\nonumber
\|Q^\blambda v\|^2=\smallsum{i\in\Z}(1{-}\lambda_i)^2|\langle \hat \pi_i,v\rangle|^2
<\smallsum{i\in\Z}|\langle \hat \pi_i,v\rangle|^2=\|v\|^2
\end{equation}
since $(1{-}\lambda_i)^2<1$ for all $i\in\Z$ and $\langle\hat\pi_{i_0},v\rangle\neq0$. As a result, \eqref{double-Pyth} implies that  $\|Q^\blambda u\|<\|u\|$.

\subsection{Proof of Proposition \ref{prop:frame-opt}}\label{app:frame-opt}

As a generalization of \eqref{op-expr}, one sees from \eqref{Mu2} that
$$M^\lambda u=u-\lambda\,S^{*\!}Su$$
for all $u\in\scB$. This shows that $M^\lambda$ is self-adjoint on $\scB$, and hence on the invariant subspace $\scV_f$. Thus, as a basic result of functional analysis \cite[\S2.13]{Conway90},
$$\|M^\lambda\|=\sup\limits_{u\in\scV_f\backslash\{0\}}\midfrac{\big|\langle u,M^\lambda u\rangle\big|}{\|u\|^2}.$$
Since
$\langle u,S^{*\!}Su\rangle=\langle Su,Su\rangle\subsmall{\scD}=\|Su\|\subsmall{\scD}^2,$
then
$\langle u,M^\lambda u\rangle=\langle u,u\rangle-\lambda\langle u,S^{*\!}S u\rangle=\|u\|^2-\lambda\|Su\|\subsmall{\scD}^2.$ After division by $\|u\|^2$, one obtains
$$\|M^\lambda\|=\max\big(1{-}\lambda A,-(1{-}\lambda B)\big).$$
This is minimized when $1{-}\lambda A=-(1{-}\lambda B)$, which gives $\lambda=2/(A{+}B)$. In this case, $\|M^\lambda\|=(B{-}A)/(B{+}A)$.

\subsection{Case of surjective operator $S$}\label{app:surjective}

The goal of this section is to show that $\ran(S)=\scD$ (and hence $S$ is surjective) when the sampling-step sequence $(T_i)\subi$ is 2-periodic with an average $T\geq1$. This is a case that trivially guarantees that $\ran(S)$ is closed, and hence that $S^\dagger$ exists. Up to a change of origin, the sequence $(t_i)\subi$ can always be put in the form of
$$t_i=\big(i+(-1)^i\delta\big)T$$
for some $\delta\in[0,\half)$. Note in this case that $T_\m=(1+2\delta)T$. The case where $T=1$ was analyzed in Appendix \ref{app:2per}. We proved there that $x(t)$ can be uniquely recovered from $\bs=(\s_i)\subi$ for any given $\bs\in\scD$. This actually implies that the operator $S$ of \eqref{S} is invertible, with $\ran(S)=\scD$ as a particular consequence. Assume now that $T>1$. Let $\scB_T$ be the subspace of $\scB$ of signals of Nyquist period $T$. The restriction $S$ to $\scB_T$ has exactly the same properties as $S$ in Appendix \ref{app:2per} where $T=1$, up to some time renormalization. So we already know that $S(\scB_T)=\scD$. Since $\scB_T\subset\scB$, we then obviously have $\ran(S)=\scD$.

\subsection{Growth control of function $h(t)$ of \eqref{h}}\label{num-trick}

\begin{app-prop}\label{prop:growth}
For any distinct $i,j\in\Z$, the inner-product $\langle f_i,f_j\rangle$ of \eqref{sksi-int} yields the alternative expression
\begin{equation}\nonumber%\label{sksi-int-alt}
\langle f_i,f_j\rangle=
\bar h(T_{i,j-1})-\bar h(T_{i-1,j-1})-\bar h(T_{i,j})+\bar h(T_{i-1,j})
\end{equation}
with any function of the type
$\bar h(t)=h(t)-\big(\alpha|t|+\beta\big)$.
\end{app-prop}
\medskip
\begin{IEEEproof}
Let $i$ and $j$ be given integers in $\Z$ and let us write $\big(T_{i,j-1},T_{i-1,j-1},T_{i,j},T_{i-1,j}\big)=(d_0,d_1,d_2,d_3)$ for convenience. When $i\neq j$, it is easy to see that $d_0,d_1,d_2,d_3$ all have the sign of $i{-}j$ (including the possibility of a 0 value). As $ a_\varphi(t)$ is an even function, it can be checked from \eqref{h} that $h(t)$ is even as well. So is $\bar h(t)$. Without loss of generality, we can then assume that $i>j$. In this case,
\begin{align*}
&\langle f_i,f_j\rangle=h(d_0)-h(d_1)-h(d_2)+h(d_3)\\
&=\bar h(d_0)-\bar h(d_1)-\bar h(d_2)+\bar h(d_3)+\alpha(d_0-d_1-d_2+d_3)
\end{align*}
where the last term is easily checked to be 0.
\end{IEEEproof}
\medskip
The growth of $\bar h(t)$ can be limited by taking $\alpha=\int_0^\infty a_\varphi(s)\dif s$ as one can show that $h(t)\sim\alpha t$ when $t$ goes to infinity. With this value of $\alpha$, it can be proved that there even exists $\beta$ such that $\bar h(t)$ vanishes at infinity, at least when $ a_\varphi(t)=O(t^{-\gamma})$ for some $\gamma>2$. This is also the case when $\varphi(t)$ is the sinc function $\sin(\pi t)/(\pi t)$ with $\alpha=\half$ and $\beta=-\frac{1}{\pi^2}$ due to the following result.
\line
\begin{app-prop}
When $\varphi(t)=\sinc(t)$, the function $h(t)$ of \eqref{h} is such that $h(t)=\frac{t}{2}-\frac{1}{\pi^2}+O(\frac{1}{t})$.
\end{app-prop}
\line
\begin{IEEEproof}
Since the Fourier transform $\Phi(\om)$ of $\varphi(t)$ is the rectangular function equal to 1 in $[-\pi,\pi]$, then $\int_0^\infty\varphi(\tau)\dif \tau=\frac{\Phi(0)}{2}=\frac{1}{2}$. Since $a_\varphi(t)=\varphi(t)$ in the present case, \eqref{h} yields
\begin{align*}
&h(t)=\textstyle t\int\limits_0^t\varphi(\tau)\dif\tau-
\int\limits_0^t \tau\varphi(\tau)\dif\tau\\
&=\textstyle t\Big(\frac{1}{2}-\int\limits_t^\infty\varphi(\tau)\dif \tau\Big)
-\int\limits_0^t\frac{\sin(\pi \tau)}{\pi}\dif\tau
=\frac{t}{2}-\frac{1}{\pi^2}+h_1(t)
\end{align*}
where
$h_1(t):=\textstyle-t\int\limits_t^\infty\varphi(\tau)\dif \tau
+\frac{\cos(\pi t)}{\pi^2}.$
By integration by parts,
\begin{eqnarray*}
&\textstyle\int\limits_t^\infty\varphi(\tau)\dif \tau
=\int\limits_t^\infty\frac{\sin(\pi \tau)}{\pi \tau}\dif \tau
=\textstyle\left[-\frac{\cos(\pi \tau)}{\pi^2 \tau}\right]_t^\infty-
\int\limits_t^\infty\frac{\cos(\pi \tau)}{\pi^2 \tau^2}\dif \tau,\\
&\textstyle\int\limits_t^\infty\frac{\cos(\pi \tau)}{\pi^2 \tau^2}\dif \tau
=\textstyle\left[\frac{\sin(\pi \tau)}{\pi^3 \tau^2}\right]_t^\infty-
\int\limits_t^\infty-2\frac{\sin(\pi \tau)}{\pi^3 \tau^3}\dif \tau
\end{eqnarray*}
so that
$\textstyle t\int\limits_t^\infty\varphi(\tau)\dif \tau=
\frac{\cos(\pi t)}{\pi^2}+\frac{\sin(\pi t)}{\pi^3 t}-
2t\int\limits_t^\infty\frac{\sin(\pi \tau)}{\pi^3 \tau^3}\dif \tau.$
Thus, $h_1(t)=-\frac{\sin(\pi t)}{\pi^3 t}+
2t\int\limits_t^\infty\frac{\sin(\pi \tau)}{\pi^3 \tau^3}\dif \tau=O(\frac{1}{t}).$
\end{IEEEproof}

\ppnoi
To calculate $\langle f_i,f_i\rangle$, however, one needs to return to the original formula \eqref{sksi-int} which yields
$\langle f_i,f_i\rangle=2 h(T_i)$
using the even symmetry of $h(t)$ mentioned in the proof of Proposition \ref{prop:growth}. This has the drawback to require a separate lookup table for $h(t)$. The length of this table however remains limited since $T_i$ remains of the order of the Nyquist period. In the system of Fig. \ref{fig:table-read}, this table would be specifically used to calculate the output coefficient $\hat\a^L_{k}=\big\langle f_{k-L},f_{k-L}\big\rangle=2 h(T_{k-L})$.

\bibliographystyle{ieeetr}

\bibliography{reference}{}

\end{document}